\documentclass[twocolumn,preprintnumbers,amsmath,amssymb,floatfix,
               aps, prc, superscriptaddress]{revtex4-1}

\usepackage{dcolumn}
\usepackage{bm}
\usepackage{multirow}
\usepackage[american]{babel}
\usepackage{color}

\def\phob         {PHOBOS}
\def\prap         {pseudorapidity}

\def\mrap         {mid-rapidity}
\def\cheren       {Cherenkov}

\newcommand {\fig}[1]{Fig.~\ref{#1}}

\newcommand {\Figs}[2]{Figures~\ref{#1} and~\ref{#2}}
\newcommand {\eq}[1]{Eq.~\ref{#1}}

\newcommand {\sect}[1]{Sect.~\ref{#1}}

\newcommand {\tab}[1]{Table~\ref{#1}}

\newcommand {\Tab}[1]{Table~\ref{#1}}

\newcommand {\snn}      {\sqrt{s_{\scriptscriptstyle{{\rm NN}}}}}

\newcommand {\dnchdeta} {\ensuremath{d\!N_{\mathrm{ch}}/\mathit{d\!\eta}}}
\newcommand {\dnchdetapA} 
            {\ensuremath{d\!N^{\mathit{pAu}}_{\mathrm{ch}}/\mathit{d\!\eta}}}
\newcommand {\dnchdetanA} 
            {\ensuremath{d\!N^{\mathit{nAu}}_{\mathrm{ch}}/\mathit{d\!\eta}}}
\newcommand {\dnchdetapp} 
            {\ensuremath{d\!N^{\mathit{p\bar{p}}}_{\mathrm{ch}}
            /\mathit{d\!\eta}}}

\newcommand {\dnhpndetapA} 
            {\ensuremath{d\!N^{\mathit{pAu}}_{h^{\pm}}/\mathit{d\!\eta}}}
\newcommand {\dnhpndetanA} 
            {\ensuremath{d\!N^{\mathit{nAu}}_{h^{\pm}}/\mathit{d\!\eta}}}
\newcommand {\RpnAu}    {\ensuremath{R_{\mathit{pnAu}}}}

\newcommand {\RdAu}     {\ensuremath{R_{\mathit{dAu}}}}
\newcommand {\abs}[1]   {\ensuremath{\left| #1 \right|}}
\newcommand {\ave}[1]   {\ensuremath{\left< #1 \right>}}
\newcommand {\prn}[1]   {\ensuremath{\left( #1 \right)}}

\newcommand {\hpos}     {\ensuremath{\mathrm{h}^+}}
\newcommand {\hneg}     {\ensuremath{\mathrm{h}^-}}
\newcommand {\have}     {\ensuremath{(\mathrm{h}^+ + \mathrm{h}^-) / 2}}

\newcommand {\pt}       {\ensuremath{p_{\mathrm{T}}}}
\newcommand {\pz}       {\ensuremath{p_{\mathrm{z}}}}

\newcommand {\mpi}      {\ensuremath{m_\pi}}

\newcommand {\Npard}    {\ensuremath{N_{\rm part}^{d}}}
\newcommand {\NparA}    {\ensuremath{N_{\rm part}^{Au}}}
\newcommand {\Ncoll}    {\ensuremath{N_{\rm coll}}}
\newcommand {\Npart}    {\ensuremath{N_{\rm part}}}

\newcommand {\aveeta}   {\ensuremath{\ave{\smash[b]{\eta}}}}
\newcommand {\avenpt}   {\ensuremath{\ave{\smash[b]{\Npart}}}}
\newcommand {\avencl}   {\ensuremath{\ave{\smash[b]{\Ncoll}}}}

\newcommand {\epcal}    {\mbox{EPcal}}

\newcommand {\pp}    {\mbox{$p$+$p$}}
\newcommand {\pbarp} {\mbox{$p$+$\bar{p}$}}
\newcommand {\Au}    {\mbox{Au}}
\newcommand {\AuAu}  {\mbox{Au+Au}}
\newcommand {\CuCu}  {\mbox{Cu+Cu}}
\newcommand {\dAu}   {\mbox{$d$+Au}}
\newcommand {\pAu}   {\mbox{$p$+Au}}
\newcommand {\nAu}   {\mbox{$n$+Au}}
\newcommand {\NAu}   {\mbox{nucleon+Au}}
\newcommand {\pA}    {\mbox{$p$+A}}

\newcommand {\dA}    {\mbox{$d$+A}}

\newcommand {\gev}   {\mbox{${\rm GeV}$}}
\newcommand {\mev}   {\mbox{${\rm MeV}$}}

\newcommand {\mom}   {\mbox{\rm GeV$\kern-0.15em /\kern-0.12em c$}}
\newcommand {\mmom}  {\mbox{\rm MeV$\kern-0.15em /\kern-0.12em c$}}
\newcommand {\mass}  {\mbox{\rm GeV$\kern-0.15em /\kern-0.12em c^2$}}
\newcommand {\mmass} {\mbox{\rm MeV$\kern-0.15em /\kern-0.12em c^2$}}

\newcommand {\cm}    {\mbox{${\rm cm}$}}

\newcommand {\fm}    {\mbox{${\rm fm}$}}

\newcommand {\us}    {\mbox{$\mu{\rm s}$}}
\newcommand {\ns}    {\mbox{${\rm ns}$}}

\newcommand {\mb}    {\mbox{${\rm mb}$}}

\newcommand {\hrefurl}[1]{\href{#1}{\url{#1}}}

\def\thetitle     {Nucleon-Gold Collisions at 200~A${\cdot}$GeV Using Tagged d+Au Interactions in {\phob}}
\def\thedate      {\today}

\ifx\pdfoutput\undefined
\else
 \pdfoutput=1
\fi
\usepackage{ifpdf}

\ifpdf
 \usepackage[pdftex]{graphicx}
 \usepackage[linkbordercolor={0 .5 0},
             citebordercolor={.8 0 0},
             urlbordercolor={0 0 .5},
             raiselinks=true,
             pdfborder={0 0 1 [2]},
             breaklinks]{hyperref}

 \pdfoutput=1        
 \pdfcatalog{/PageMode(/UseNone)}
\else 

 \usepackage[dvips]{graphicx}
  \usepackage[linkbordercolor={0 .5 0},
              citebordercolor={.8 0 0},
              urlbordercolor={0 0 .5},
              raiselinks=true,
              breaklinks]{hyperref}
 \usepackage{breakurl}
\fi

\graphicspath{ {./img/} }
\begin{document}

\title{\thetitle}
%
%

\def\ANL  {Argonne National Laboratory, Argonne, IL 60439-4843, USA}
\def\BNL  {Brookhaven National Laboratory, Upton, NY 11973-5000, USA}
\def\INP  {Institute of Nuclear Physics, Krak\'{o}w, Poland}
\def\MIT  {Massachusetts Institute of Technology, Cambridge, MA 02139-4307, USA}
\def\NCU  {National Central University, Chung-Li, Taiwan}
\def\UIC  {University of Illinois at Chicago, Chicago, IL 60607-7059, USA}
\def\MCP  {University of Maryland, College Park, MD 20742, USA}
\def\URR  {University of Rochester, Rochester, NY 14627, USA}

\author{B.B.Back}\affiliation{\ANL}
\author{M.D.Baker}\affiliation{\BNL}
\author{M.Ballintijn}\affiliation{\MIT}
\author{D.S.Barton}\affiliation{\BNL}
\author{B.Becker}\affiliation{\BNL}
\author{R.R.Betts}\affiliation{\UIC}
\author{A.A.Bickley}\affiliation{\MCP}
\author{R.Bindel}\affiliation{\MCP}
\author{W.Busza}\affiliation{\MIT}
\author{A.Carroll}\affiliation{\BNL}
\author{M.P.Decowski}\affiliation{\MIT}
\author{E.Garc\'{\i}a}\affiliation{\UIC}
\author{T.Gburek}\affiliation{\INP}
\author{N.George}\affiliation{\BNL}
\author{K.Gulbrandsen}\affiliation{\MIT}
\author{S.Gushue}\affiliation{\BNL}
\author{C.Halliwell}\affiliation{\UIC}
\author{J.Hamblen}\affiliation{\URR}
\author{A.S.Harrington}\affiliation{\URR}
\author{C.Henderson}\affiliation{\MIT}
\author{D.J.Hofman}\affiliation{\UIC}
\author{R.S.Hollis}\affiliation{\UIC}
\author{R.Ho\l y\'{n}ski}\affiliation{\INP}
\author{B.Holzman}\affiliation{\BNL}
\author{A.Iordanova}\affiliation{\UIC}
\author{E.Johnson}\affiliation{\URR}
\author{J.L.Kane}\affiliation{\MIT}
\author{N.Khan}\affiliation{\URR}
\author{P.Kulinich}\affiliation{\MIT}
\author{C.M.Kuo}\affiliation{\NCU}
\author{J.W.Lee}\affiliation{\MIT}
\author{W.T.Lin}\affiliation{\NCU}
\author{S.Manly}\affiliation{\URR}
\author{A.C.Mignerey}\affiliation{\MCP}
\author{R.Nouicer}\affiliation{\BNL}\affiliation{\UIC}
\author{A.Olszewski}\affiliation{\INP}
\author{R.Pak}\affiliation{\BNL}
\author{I.C.Park}\affiliation{\URR}
\author{H.Pernegger}\affiliation{\MIT}
\author{C.Reed}\affiliation{\MIT}
\author{C.Roland}\affiliation{\MIT}
\author{G.Roland}\affiliation{\MIT}
\author{J.Sagerer}\affiliation{\UIC}
\author{P.Sarin}\affiliation{\MIT}
\author{I.Sedykh}\affiliation{\BNL}
\author{W.Skulski}\affiliation{\URR}
\author{C.E.Smith}\affiliation{\UIC}
\author{P.Steinberg}\affiliation{\BNL}
\author{G.S.F.Stephans}\affiliation{\MIT}
\author{A.Sukhanov}\affiliation{\BNL}
\author{M.B.Tonjes}\affiliation{\MCP}
\author{A.Trzupek}\affiliation{\INP}
\author{C.Vale}\affiliation{\MIT}
\author{G.J.\surname{van~Nieuwenhuizen}}\affiliation{\MIT}
\author{R.Verdier}\affiliation{\MIT}
\author{G.I.Veres}\affiliation{\MIT}
\author{F.L.H.Wolfs}\affiliation{\URR}
\author{B.Wosiek}\affiliation{\INP}
\author{K.Wo\'{z}niak}\affiliation{\INP}
\author{B.Wys\l ouch}\affiliation{\MIT}
\author{J.Zhang}\affiliation{\MIT}

\collaboration{PHOBOS Collaboration}

\date{\thedate}

\begin{abstract}\noindent

Forward calorimetry in the {\phob} detector has been used to study
charged hadron production in {\dAu}, {\pAu} and {\nAu} collisions at
$\snn = 200~{\gev}$. The forward proton calorimeter detectors are
described and a procedure for determining collision centrality with
these detectors is detailed. The deposition of energy by deuteron
spectator nucleons in the forward calorimeters is used to identify
{\pAu} and {\nAu} collisions in the data. A weighted combination of
the yield of {\pAu} and {\nAu} is constructed to build a reference for
{\AuAu} collisions that better matches the isospin composition of the
gold nucleus. The  $\pt$ and centrality dependence of the yield of this improved reference
system is found to match that of {\dAu}. The
shape of the charged particle transverse momentum distribution is
observed to extrapolate smoothly from {\pbarp} to central {\dAu} as a
function of the charged particle {\prap} density. The asymmetry of
positively- and negatively-charged hadron production in {\pAu} is compared
to that of {\nAu}. No significant asymmetry is observed at
{\mrap}. These studies augment recent results from experiments at the
LHC and RHIC facilities to give a more complete description of
particle production in {\pA} and {\dA} collisions, essential for the
understanding the medium produced in high energy nucleus-nucleus
collisions.

\vspace{3mm}
\end{abstract}

\maketitle

\section{Introduction}\label{intro}


The {\phob} detector~\cite{Back:2003sr} at the Relativistic Heavy Ion
Collider (RHIC)~\cite{Hahn:2003sc} is one of several
experiments~\cite{Adamczyk:2003sq, Adcox:2003zm, Ackermann:2002ad}
that have measured the invariant yield of charged hadrons in
collisions of deuterons with gold nuclei at a nucleon-nucleon center
of mass energy of $\snn = 200~{\gev}$.  In the referenced papers, charged hadron production is
studied as a function of both transverse momentum ($\pt$) and
collision centrality (a measure correlated with the impact parameter
of the deuteron). The particle yields for $\pt$ above about
1.5--2.0~{\mom} are similar to, or possibly slightly enhanced above,
those observed in {\pbarp} collisions at the same
energy~\cite{Back:2003ns}, somewhat reminiscent of the so-called
Cronin effect seen in proton-nucleus collisions~\cite{Cronin:1974zm}.
Previous analyses of the {\dAu} charged hadron spectra by
{\phob}~\cite{Back:2003ns} and the other RHIC
experiments~\cite*{Adler:2003ii, Adams:2003im, Arsene:2003yk,
  Adams:2006nd, Adler:2006xd} have demonstrated that this enhancement
stands in stark contrast to the observed suppression of high $\pt$
hadrons in the (central) {\AuAu} collision system at $\snn =
200~{\gev}$~\cite*{Adcox:2001jp, Adcox:2002pe, Adler:2002xw,
  Back:2003qr}. Since no suppression is found in {\dAu} collisions,
the effect seen in central {\AuAu} interactions has been interpreted
as evidence of final state effects, in particular parton energy loss.
It should be noted that evidence of possible collective effects in
systems such has {\dAu} and $p$+Pb have been found recently, but
only for events with very high final state particle multiplicity (see,
as one example, Ref.~\cite{pPbFlow}).

The choice of the reference system used in comparing to {\AuAu} data,
and of the centrality measure, are both of critical importance to the
understanding of the observed suppression. The data and Monte Carlo
(MC) simulations presented in this paper are used to study the
choice of centrality measure, as well as the choice of reference
system. Centrality measures based on the multiplicity of particles in
the high-{\prap} region as well as on the number of spectators in the
gold nucleus are examined.
To study the chosen reference system, a calorimetry-based technique is
used to identify, on an event-by-event basis, the subsets of {\dAu}
collisions in which only the proton or only the neutron participated
in the collision.  
Specifically, a calorimeter on the side of the interaction region
where the Au beam exits is used
as part of the determination of collision centrality while a second
calorimeter on the other side is used in the selection of
{\nAu} and {\pAu} interactions.
Similar tagging of the nucleon+Au component of the
{\dAu} data has also been investigated by the PHENIX
collaboration~\cite{PHENIX_Tag, PHENIX_Tag2}.  These nucleon-nucleus collisions are
used to construct an ideal reference system for comparison with
{\AuAu} collisions.  Further, the charged hadron yields of {\nAu} and
{\pAu} are compared in order to study the ability of nucleon-nucleus
collisions to transport charge to the {\mrap} region.

\section{The {\phob} Detector}
\label{intro:detector}

The {\phob} experiment makes use of multiple detector components to
measure particles produced by collisions at RHIC. Silicon pad
detectors near the interaction point are used for particle tracking
and collision vertex determination, see \sect{spec}. Additional
silicon pad detectors provide full azimuth and large {\prap} coverage,
as described in \sect{coll}. Collision triggering is provided by
plastic scintillator arrays at high {\prap}, see \sect{coll}, and by
calorimeters measuring the number of neutral spectator nucleons,
described below. More detail on these subsystems may be found in
Ref.~\cite{Back:2003sr}.


To study nucleon-nucleus collisions, two calorimeters were added to
the {\phob} experiment prior to the 2003 {\dAu} physics run at RHIC.
These detectors extend the measurement of forward-going nuclear
fragments. Complementing the pre-existing zero-degree calorimeters
(``ZDCs'') that collect energy from spectator
neutrons~\cite{Adler:2003sp}, the proton calorimeter (``PCAL'')
detectors measure energy from free spectator protons. Each PCAL
detector is assembled from lead-scintillator bricks originally
constructed for the E864 experiment~\cite{Armstrong:1998qs} at the
AGS. The bricks are 117~{\cm} in length with a $10\times10$~{\cm}
cross section facing the interaction point. Each brick has an array of
$47\times47$ scintillating fibers running along its entire length. All
of the fibers from a single detector element are read out by a
Phillips XP 2262B phototube at the back.


\begin{figure}[t]
   \centering
   \includegraphics[width=\linewidth]{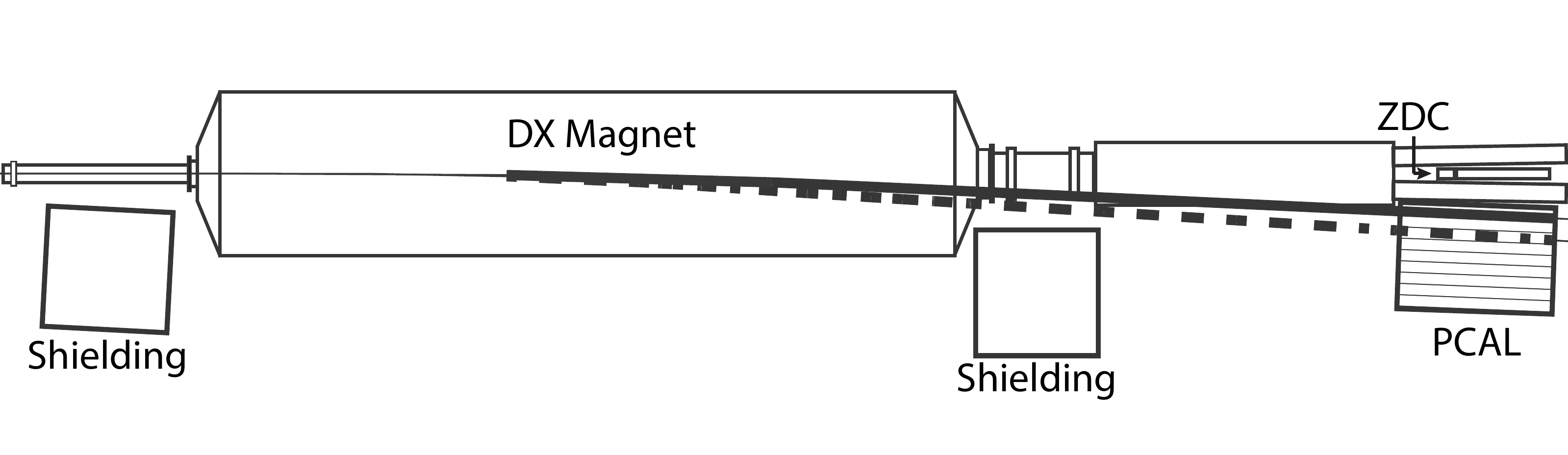}
   \includegraphics[width=\linewidth]{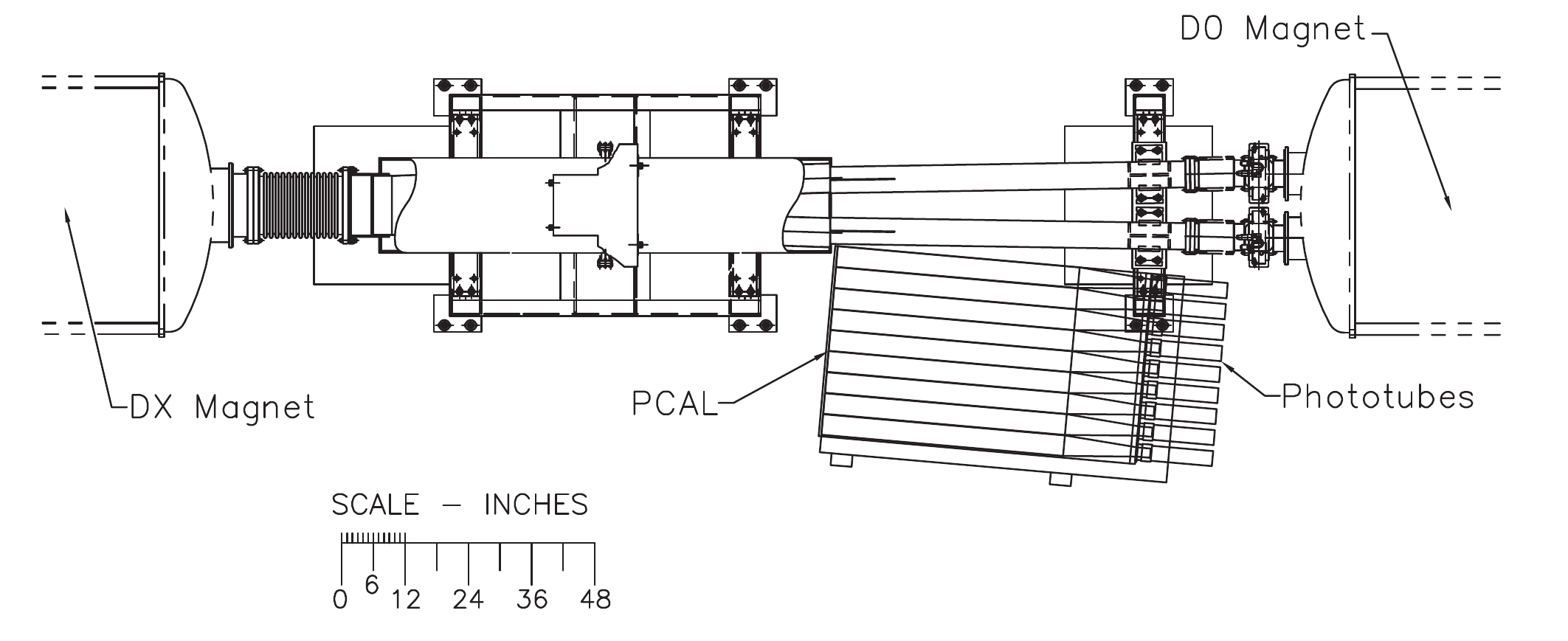}
   \caption{\label{exp:fig:pcal}Top:  Schematic overview of the {\phob} {\Au}-PCAL, also showing the shielding and  ZDC. The solid (dashed) lines show the approximate trajectories followed by spectator protons from the {\Au}
     nucleus with momenta of 100 (50)~GeV/$c$ as they are bent by the DX magnet
     into the calorimeter. Bottom: Detailed expanded view of the PCAL region (right half of the upper figure), including the DX and D0 accelerator magnets. The scale shown applies only to this detailed view. The  shielding and  ZDC detector
     are not shown in the bottom image.}
\end{figure}

The PCAL detector on the {\Au}-exit side of the collision (see plan
view in \fig{exp:fig:pcal}) consists of
an array 8~bricks wide by 10~bricks high. The d-exit side PCAL (not
shown in the figure) is a small $2\times2$ array. As mentioned above,
the former is used for centrality determination while the latter is
used, along with the ZDC, for tagging {\nAu} and {\pAu} interactions.
Both calorimeters are centered at the beam height and the smaller
calorimeter is mounted with its elements at the same location transverse 
to the beam as the two closest elements shown in \fig{exp:fig:pcal}.

Because of their higher charge to mass ratio
(compared to the deuteron and Au nuclei, as well as nuclear fragments), spectator protons
emerging from either side of the interaction are bent out of the beam
pipe and into a PCAL detector by the RHIC DX-magnets.  The primary
purpose of these DX-magnets is to direct the deuteron and gold ion
beams into and out of the interaction region.

The larger {\Au}-PCAL covers a {\prap} region $-3.6<\eta<-5.2$ and
therefore could be struck by produced particles in addition to the
spectator protons it was designed to detect. In order to prevent this,
two shields consisting of 44~{\cm} thick concrete blocks were
installed between the calorimeter and the interaction region. 


The energy coming from {\Au}-side spectator protons ({\epcal}) is
calculated using bricks in the the {\Au}-PCAL which are located in the
two rows at beam height, as well as the outer four columns away from
the beam. The two rows at beam height are found to contain a majority
of the hadronic shower energy in simulations of single nucleons having
momenta comparable to nuclear thermal and fragmentation emission. The
columns away from the beam supplement the shower containment. The
remaining bricks, in columns near the beam but above and below beam
height, are not included in {\epcal}. This reduces contamination from
particles emitted in the neutron-induced hadronic showers which escape
the ZDC. 

The {\Au}-PCAL modules have been calibrated relative to each other
using energy deposited by cosmic rays. Fast scintillator detectors are
installed above and below the {\Au}-PCAL detector to serve as cosmic
ray triggers during dedicated calibration data taking. Modules in the
d-PCAL have been calibrated relative to each other by minimizing the
width of the single-proton peak in the d-PCAL energy distribution.

\section{Collision Reconstruction}\label{coll}

\subsection{Collision Selection}\label{coll:trig}

Deuteron-gold interactions are identified using a set of selection
criteria designed to minimize background (i.e.~beam-gas interactions)
and enhance the sample of collisions which could produce particles
inside the spectrometer acceptance. First, at least one hit is
required in both of the 16-scintillator arrays (see
Ref.~\cite{Back:2003sr} for more details on this and other detector elements) which
cover a {\prap} range of $3 < \abs{\eta} < 4.5$. Then, the
longitudinal collision vertex, as determined by a single-layer silicon
detector covering the beam-pipe in the {\mrap} range, is
required to be within 10~{\cm} of the nominal interaction point.
Further, this vertex is required to be in reasonable agreement
(within 25~{\cm}) with that found by the simple timing difference of two sets
of fast {\cheren} counters, located at $-4.9 < \eta < -4.4$ and $3.6 <
\eta < 4.1$ ($\eta > 0$ being in the deuteron direction). Finally,
events that appeared to have signals from either a previous or
following collision are removed. If two events occur within 5~{\us},
the later event is rejected as containing pile-up signals in the
silicon. If two events occur within 500~{\ns}, as determined using the
fast trigger detectors, then both are rejected as pile-up.

\subsection{Centrality Determination}\label{coll:cent}

Two experimental observables have been used as centrality measures by
the analysis presented in this paper. The first variable, ``ERing'',
is a measure of the total energy recorded in ``Rings'', endcap silicon
detectors. The rings have nearly $2\pi$ coverage in azimuth and cover
eta ranges of $-5.4<\eta<-3.0$ and $3.0<\eta<5.4$. The second
variable, {\epcal}, is described in \sect{intro:detector}, and measures
the energy of {\Au} protons that do not participate in inelastic
collisions with the deuteron. Thus, {\epcal} measures protons near beam
rapidity, $y=5.36$.



The distribution of each of these variables in the {\dAu} data can be
used to determine the fractional cross section centrality
bins. Details on this procedure may be found in
Refs.~\cite{Back:2003hx,Back:2004mr}.  The extraction of average
values of collision parameters, such as the number of participant
nucleons ({\Npart}), as well as the determination of the
centrality-dependent efficiency of the collision event selection
requires a set of simulations. Models of {\dAu} collisions from both
the HIJING~\cite{Gyulassy:1994ew} and AMPT~\cite{Zhang:1999bd}
packages have been studied. The detector simulation has been performed
using the GEANT package~\cite{Geant}. In addition to {\Npart}, other
centrality parameters have been studied using these simulations,
including $\NparA$ and $\Npard$, the number of participants in the
gold and deuteron, respectively, $\Ncoll$, the number of
nucleon-nucleon collisions in the interaction, and $\nu$, the average
number of collisions per deuteron participant.

\begin{figure}[t]
   \begin{center}
      \includegraphics[width=0.95\linewidth]{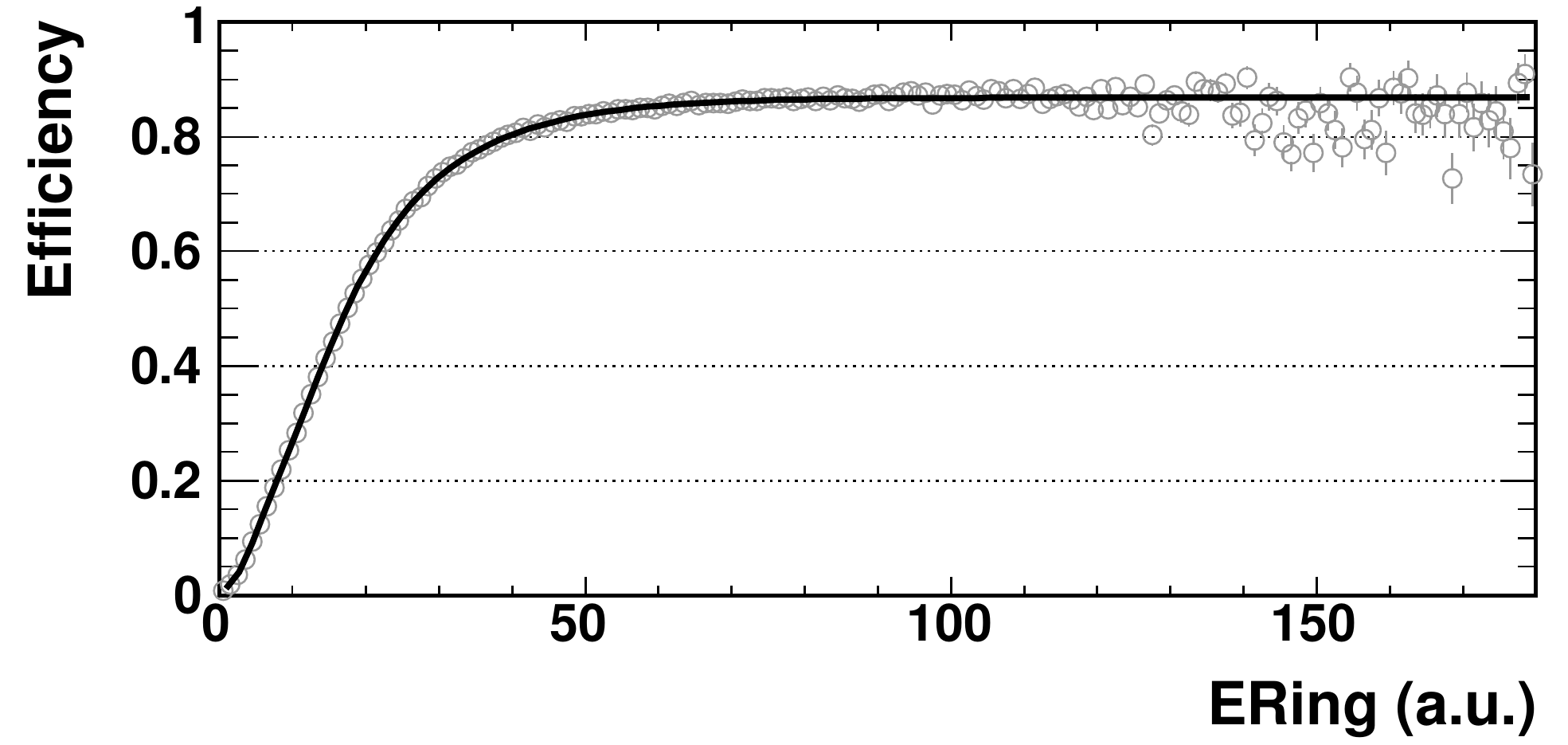}
   \end{center}
   \caption{\label{ana:fig:ERingEfficiencies}
     The event selection efficiency as a function of the ERing
     centrality variable. Grey points show the fraction of events
     simulated using AMPT that pass the event selection (see text for
     details). The black line represents a smooth fit to the points.}
\end{figure}


The efficiency of the collision selection can be determined from the
simulations as a function of the chosen centrality variable. This is
done by counting the fraction of simulated events that pass the event
selection as a function of centrality. Because the event selection
contains a vertex cut, the fraction is calculated as the number of
events passing the event selection divided by the number of events
having a true interaction vertex within 10~{\cm} of the nominal
interaction point. The efficiency as a function of the ERing
centrality variable, obtained using AMPT simulations, is shown in
\fig{ana:fig:ERingEfficiencies}. Note that the efficiency does not
approach unity, even for central events, due to the small acceptance
of the detectors that determine the collision vertex as part of the
trigger.

This efficiency is used to unbias the centrality variable distribution
measured in the data. The unbiased distribution is then divided into
fractional cross section bins, using the method described in
Ref.~\cite{Back:2002uc}.

The efficiency function presented in \fig{ana:fig:ERingEfficiencies}
is also used to correct, on an event-by-event basis, the measurements
of the charged hadron spectra presented in this article. This accounts
for the variation of the selection efficiency within a centrality bin,
whereas the application of the average efficiency in a centrality bin
would not.

For both the HIJING and AMPT collision generators, a Glauber model has
been used to determine the average values of centrality parameters,
such as {\Npart}, which cannot be measured directly. A Hulth\'en wave
function~\cite{Hulthen} has been used to model the deuteron profile,
while the gold nucleus density has been modeled using a Woods-Saxon
distribution. The value of the inelastic nucleon-nucleon cross section used in
the Glauber model is 41~{\mb}. The average value of the chosen
centrality parameter can then be determined for each fractional cross
section bin; for details on this procedure, see
Ref.~\cite{Back:2001xy}.

The systematic uncertainties of the various (unbiased) centrality
parameters, such as {\Npart}, have been studied. The dependence on
simulations has been quantified by varying the centrality efficiency,
for example, that shown in \fig{ana:fig:ERingEfficiencies} for ERing
centrality bins. The amount by which the efficiency can vary is
estimated by dividing the simulated events into vertex bins. The
dependence on the deuteron wave function has been studied by using
both a Hulth\'en wave function, as well as a Woods-Saxon
distribution. The uncertainty of the centrality parameters resulting
from the choice of collision simulation model has been studied by
comparing to simple Glauber MCs. Uncertainties in using the efficiency
function to unbias the centrality parameters have been accounted for
by smearing the centrality measure (i.e.\ ERing) prior to applying the
efficiency correction. Finally, the centrality parameters coming from
different collision simulation packages are compared.


The centrality parameters determined from ERing centrality bins are
presented in \Tab{raa:eringCentPars}. The values for {\pAu} and {\nAu}
tagged events, described in \sect{coll:tag}, are also shown. The
systematic uncertainties of {\avenpt} and {\avencl} are typically
slightly different, with that for {\avencl} usually larger. The table
lists the larger of the two uncertainties.

\subsection{Proton Calorimeter Centrality Determination}
\label{pcal_cent}

\begin{figure}[t]
   \centering
   \includegraphics[width=0.95\linewidth]{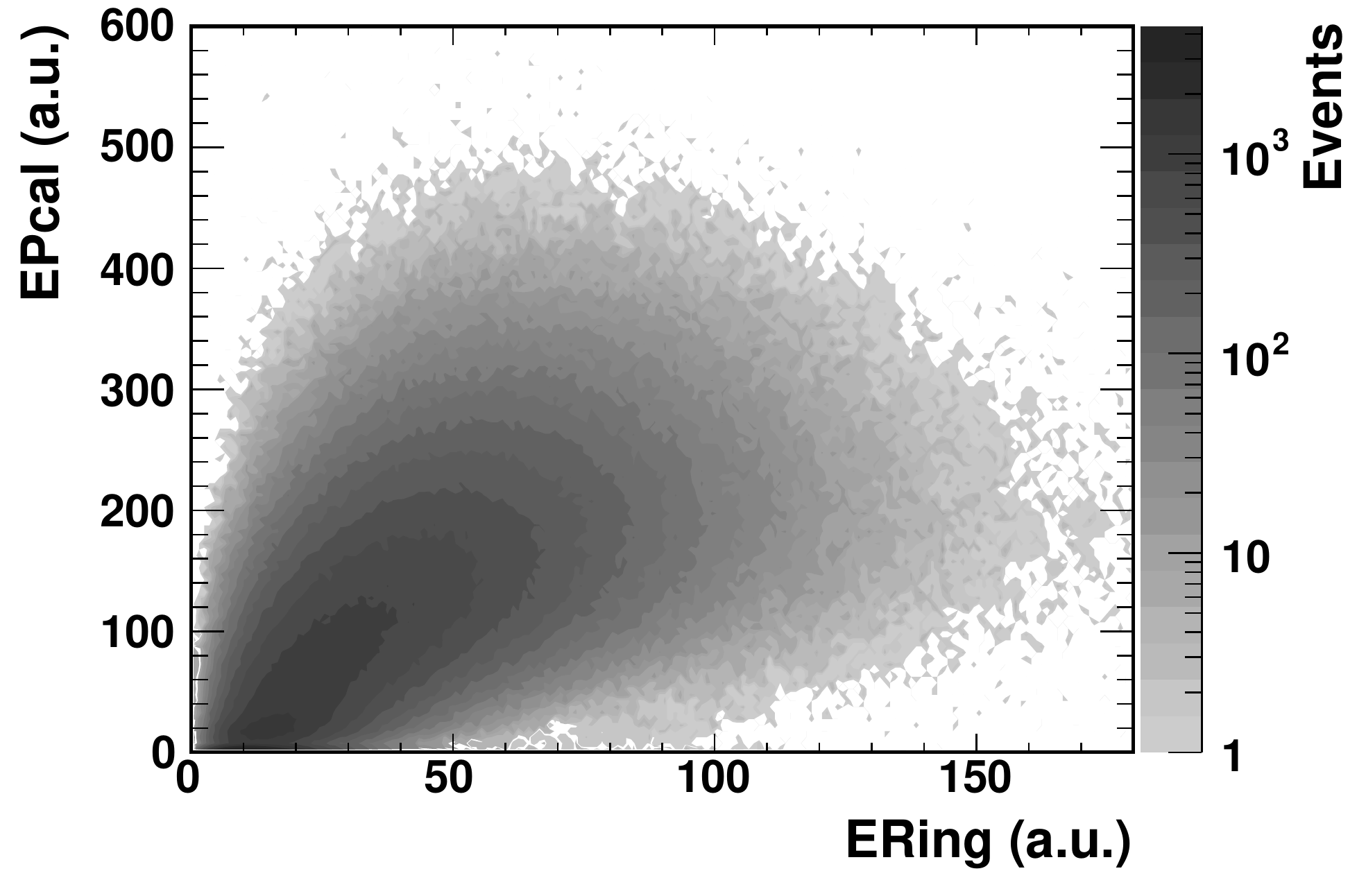}
   \caption{   \label{recon:fig:pcalvsering}
     The correlation between {\epcal} and ERing used to obtain
     {\epcal} centrality bins.
   }
\end{figure}

The Au-PCAL detector facilitates the determination of the centrality
of {\dAu} collisions using a variable, {\epcal}, which is independent of
the measured multiplicity. As has been shown
previously~\cite{Back:2003hx,Back:2004mr}, multiplicity measurements
in a particular region of {\prap} may be biased if the centrality of
collisions is determined using (multiplicity based) observables in a
similar {\prap} region. The ERing observable is measured at high
{\prap}, allowing measurements at {\mrap} to be minimally biased by
such auto-correlations. Centrality derived from the number of
spectator nucleons should be free of such biases. A measurement of the
charged hadron spectral shape in centrality bins from both ERing and
{\epcal} is presented in \sect{cronin}.

Centrality bins could, in principle, be derived from {\epcal} signals
using the same procedure as for the other observables. However, the
breakup of the gold nucleus is not modeled by either the MC event
generators, HIJING and AMPT, or the GEANT detector simulation.  As a
result, an alternative procedure has been developed that exploits the
monotonic correlation in the {\dAu} data between the {\epcal} signal, and
the signal of another (well-modeled) detector, ERing. This correlation
is shown in \fig{recon:fig:pcalvsering}.

\begin{figure}[t]
   \centering
   \includegraphics[width=0.95\linewidth]{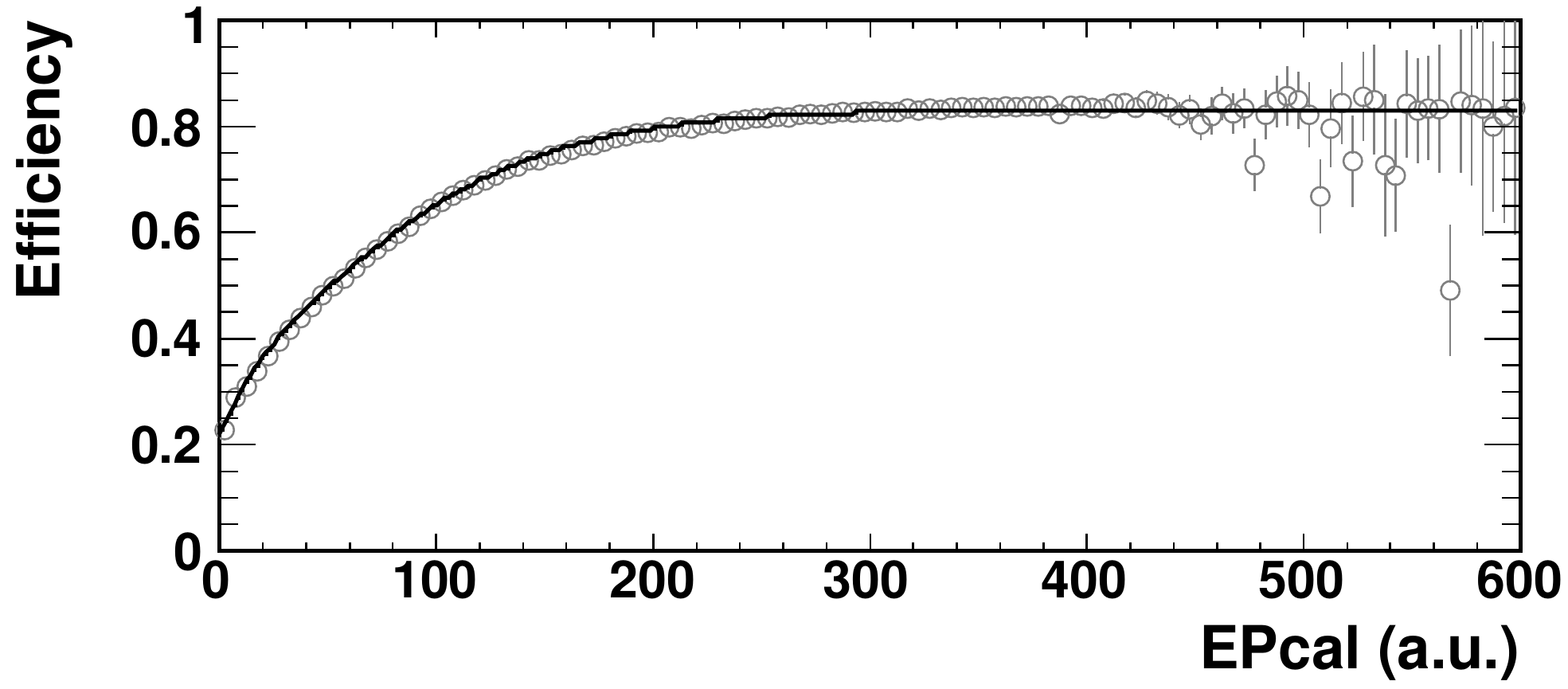}
   \caption{ \label{recon:fig:pcalEringEff}
     The event selection efficiency as a function of the {\epcal}
     centrality variable. Points represent the ratio between the
     number of events in an EPcal bin and the number of events
     expected for a perfectly efficient detector, obtained using the
     ERing efficiency function (see text for details). The black
     line is a smooth fit to the points.}
\end{figure}

\begin{figure}[t]
   \centering
   \includegraphics[width=0.95\linewidth]{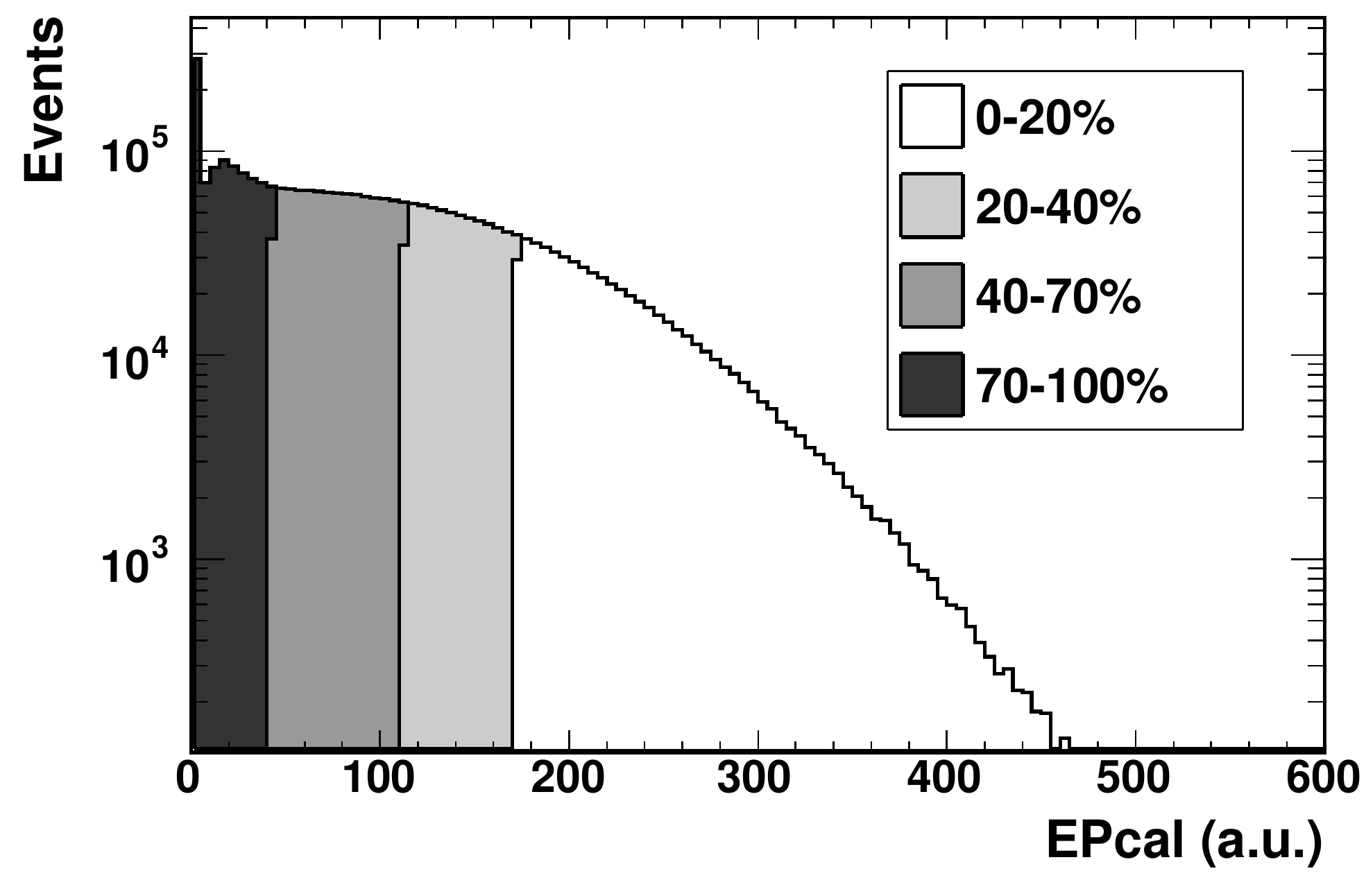}
   \caption{ \label{recon:fig:pcalEringCuts}
     The centrality bins obtained using the known ERing efficiency.
     Each slice of the histogram shows the distribution of {\epcal}
     within the specified fractional cross section bin. The kinks in
     the shaded histograms arise from the edge of a fractional cross
     section bin falling inside a histogram bin.}
\end{figure}

The method for deriving event selection efficiency for a given value
of {\epcal} uses the known efficiency of ERing. Using each event in the
data, two distributions of {\epcal} are generated: one simply counting
events and one counting events but weighted by the inverse of the
known efficiency of the correlated observable, $1/\epsilon_{ERing}$.
The efficiency as a function of {\epcal} is determined from the ratio
of the simple-count distribution divided by that using weighted
counts. This efficiency is used in the standard procedure to evaluate
{\epcal} cutoff values for the centrality bins.
\Figs{recon:fig:pcalEringEff}{recon:fig:pcalEringCuts} show the event
selection efficiency as a function of {\epcal} and the resulting
{\epcal} centrality bins, respectively, obtained by using ERing.

Two different procedures have been developed to estimate the average
number of nucleons participating in the inelastic collision, $\Npart$,
for a given {\epcal} centrality bin. Both procedures exploit the
correlation of {\epcal} with ERing and then of ERing with $\Npart$. The
same procedures are used to estimate other collision parameters as
well, such as the number of nucleon-nucleon collisions, $\Ncoll$, or
the impact parameter, $b$.

\begin{figure}[t]
   \centering
   \includegraphics[width=0.95\linewidth]{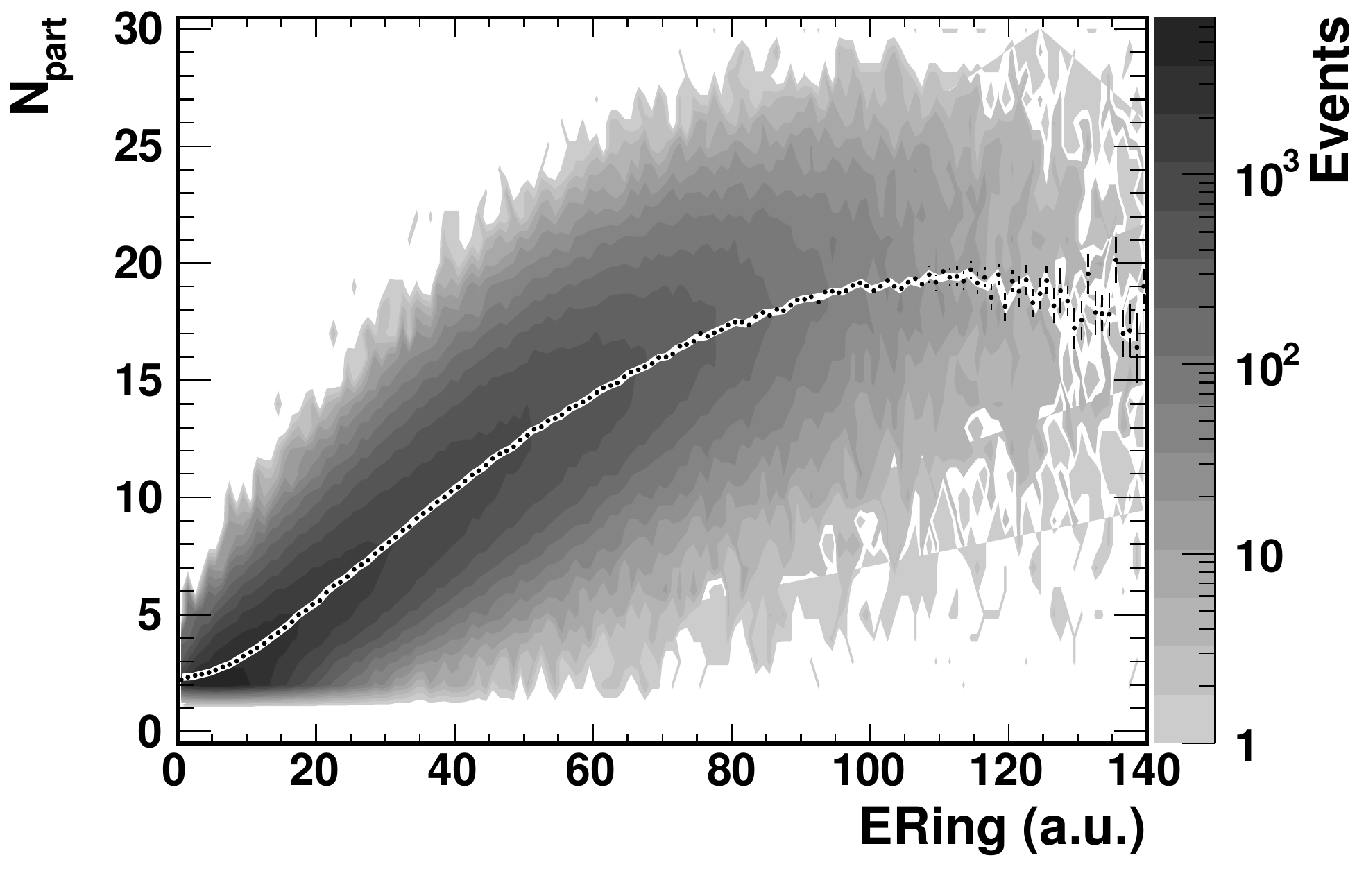}
   \caption{    \label{recon:fig:npartVsERingProf}
     $\Npart$ dependence on ERing in the MC. The white line shows the
     fit to the mean $\Npart$ in each ERing bin.}
\end{figure}

\begin{figure}[t]
   \centering
   \includegraphics[width=0.95\linewidth]{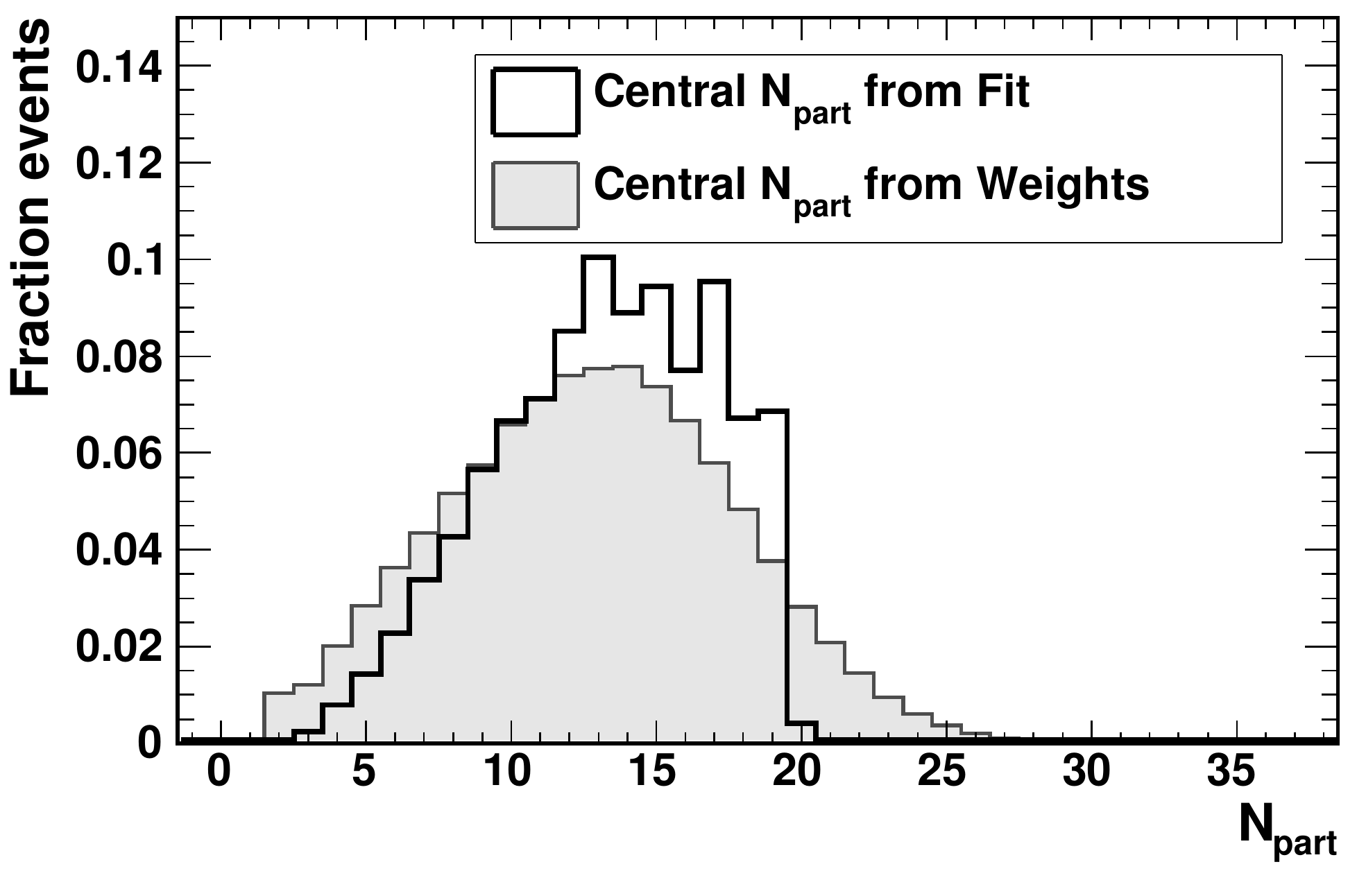}
   \caption{   \label{recon:fig:npartERingWtFitComp}
     The $\Npart$ distribution in the 0-20\% central {\epcal} bin
     found using the $\Npart$ vs ERing fit method (open histogram)
     compared to that from the weighting method (grey histogram). Each
     distribution is (independently) normalized.}
\end{figure}

The simpler approach involves fitting the mean $\Npart$ in small bins
of ERing, as shown in \fig{recon:fig:npartVsERingProf}. The fit is
used to estimate the average value of $\Npart$ given the value of
ERing in an event. These values are then used to obtain $\Npart$
distributions for each {\epcal} centrality bin.

The second approach begins by dividing the ERing distribution for
events in a given {\epcal} centrality bin by the distribution for all
events in order to determine the probability of any particular value
of ERing in that bin. Then, for each centrality bin, all MC events are
weighted according to the appropriate probability for their value of
ERing and the distribution of $\Npart$ is determined with these
weights applied.

The results of the two techniques are compared for the most central
{\epcal} bin in \fig{recon:fig:npartERingWtFitComp}. In the first
procedure (open histogram), the spread of $\Npart$ in the resulting
distribution depends only on the width of the correlation of {\epcal}
and ERing, while in the second (grey histogram) it is also affected by
the correlation of ERing and $\Npart$. The latter is almost certainly
an overestimate of the width of $\Npart$ in a given centrality bin,
while the former may underestimate the spread. However, in the
analysis of spectra and yields, this difference in width is only
significant to the degree that it affects the mean value. The
differences of the means found using the two techniques are included
in the systematic uncertainty estimate for the values of
$\Npart$. Analogous systematic uncertainties are determined for the
other centrality parameters, such as $\Ncoll$ or $b$. The weighting
and fit procedures differ by about 5\% in central {\dAu} and about
25\% in peripheral {\pAu}.

%
%

\begin{table*}
   \begin{center}
      \begin{tabular}{|r|r|l|l|l|l|l|l|l|l|}
\hline
\hline
Parameter & System(s) & \multicolumn{4}{c|}{ERing Bins} &
\multicolumn{4}{c|}{{\epcal} (from ERing) Bins} \\
 & & 0-20\% & 20-40\% & 40-70\% & 70-100\% & 0-20\% & 20-40\% & 40-70\% & 70-100\%\\ 
\hline
\multirow{2}{*}{\ave{b} ({\fm})} & {\dAu} & $3.3(1.4)$ & $4.7(1.5)$ & $6.3(1.4)$ & $7.6(1.3)$ & $4.1(1.8)$ & $4.9(2.0)$  & $6.0(1.9)$ & $7.3(1.6)$ \\
 & {\pAu}, {\nAu} & $6.1(1.4)$ & $6.4(1.3)$ & $7.2(1.3)$ & $8.0(1.3)$ & $6.9(1.4)$ & $7.2(1.4)$ & $7.6(1.4)$ & $7.9(1.3)$ \\
\hline
\multirow{2}{*}{\avenpt} & {\dAu} & $15.4(3.8)$ & $10.6(2.9)$ & $6.3(2.4)$ & $3.1(1.3)$ & $12.8(4.9)$ & $10.4(4.9)$ & $7.4(4.3)$ & $4.1(2.5)$ \\
 & {\pAu}, {\nAu} & $9.4(3.4)$ & $7.7(2.5)$ & $4.7(1.9)$ & $2.7(1.0)$ & $5.8(3.0)$ & $4.8(2.6)$ & $3.9(2.1)$ & $3.0(1.4)$ \\
\hline
\multirow{2}{*}{\avencl} & {\dAu} & $14.5(4.2)$ & $9.4(3.3)$ & $5.0(2.5)$ & $2.0(1.2)$ & $11.8(5.2)$ & $9.3(5.2)$ & $6.1(4.5)$ & $3.0(2.5)$ \\
 & {\pAu}, {\nAu} & $8.4(3.4)$ & $6.7(2.5)$ & $3.7(1.9)$ & $1.7(1.0)$ & $4.8(3.0)$ & $3.8(2.6)$ & $2.9(2.1)$ & $2.0(1.4)$ \\
\hline
\multirow{2}{*}{\ave{\nu}} & {\dAu} & $7.6(2.1)$ & $5.2(1.8)$ & $3.3(1.5)$ & $1.7(0.9)$ & $6.3(2.6)$ & $5.2(2.5)$ & $3.7(2.3)$ & $2.2(1.4)$ \\
 & {\pAu}, {\nAu} & $8.4(3.4)$ & $6.7(2.5)$ & $3.7(1.9)$ & $1.7(1.0)$ & $4.8(3.0)$ & $3.8(2.6)$ & $2.9(2.1)$ & $2.0(1.4)$ \\
\hline
\multirow{2}{*}{Sys. Error} & {\dAu} & 7.5\% & 10\% & 15\% & 30\% & 15\% & 15\% & 20\% & 30\% \\
 & {\pAu}, {\nAu} & 10\% & 12\% & 17\% & 31\% & 31\% & 31\% & 31\% & 31\% \\
\hline
\hline
      \end{tabular}
   \end{center}
    \caption{   \label{raa:eringCentPars}
       Centrality parameters determined using ERing- and
       {\epcal}-based centrality bins and AMPT collision
       simulations. Centrality bins represent the fraction of the
       total {\dAu} cross section, even for the {\pAu} and {\nAu}
       collision systems (see \sect{coll:tag:cent}). Values in
       parentheses are the RMS of their respective parameters. For the
       {\epcal} bins, the weighted ERing method has been used (see
       \sect{pcal_cent}). {\ave{b}} is the average impact parameter,
            {\avenpt} is the average number of participant nucleons,
            {\avencl} is the average number of collisions and
            {\ave{\nu}} is the average number of collisions per
            deuteron participant. The last row lists systematic
            uncertainties in {\avenpt} and {\avencl}. See text for
            discussion.}
\end{table*}

The systematic uncertainty inherent in the procedure used to determine
centrality from the {\epcal} variable has been studied. This has been
done by applying the indirect procedure described above for {\epcal} to well modeled detectors at
mid-rapidity, for which the direct procedure described in \sect{coll:cent}
can also be used. Discrepancies between centrality parameters obtained via
the direct and indirect methods are used to quantify the systematic
uncertainties on this procedure. This uncertainty is in addition to
those described in \sect{coll:cent}.

The centrality parameters found in the {\epcal} centrality bins are
presented in \Tab{raa:eringCentPars}. The parameters have been
determined using the weighting method. The table also lists the values
for {\pAu} and {\nAu} tagged events, which are described in
\sect{coll:tag}.

\subsection{Deuteron-Nucleon Tagging}\label{coll:tag}

The low binding energy of the deuteron nucleus (1.11~{\mev} per
nucleon) facilitates the analysis presented in this paper.  Because
the deuteron is so weakly bound, it is possible for the nucleons to be
relatively far apart at the moment the deuteron collides with the gold
nucleus. This can result in only one nucleon of the deuteron
participating in the (inelastic hadronic) collision. Furthermore, the
binding energy is so small compared to the beam energy that the
remaining spectator nucleon can emerge from the collision almost
completely unperturbed.  Thus, such a collision can be treated as an
effective collision between a single nucleon and a gold nucleus.

Although the size of a deuteron is relatively large, the
proton-neutron separation is typically not larger than the size of the
Au target. As a result, the nucleon-gold collisions that form a subset
of the deuteron-gold data are not equivalent to minimum bias
nucleon-gold data. Rather, they are biased towards more peripheral
interactions. Further investigations of this bias and the techniques
used to address it in the present analysis are discussed below.

The subset of {\dAu} collisions matching {\pAu} and {\nAu}
interactions have been identified through the observation of the
spectator nucleon of the deuteron. The deuteron spectators are
measured in {\phob} using both the PCAL and ZDC detectors on the
deuteron-exit side of the collision.  Qualitatively, a collision in
which the d-PCAL registered a spectator and the d-ZDC did not is
labeled an {\nAu} interaction (and vice-versa for {\pAu}
interactions).


\begin{figure}[th]
   \centering
     \includegraphics[width=0.95\linewidth]{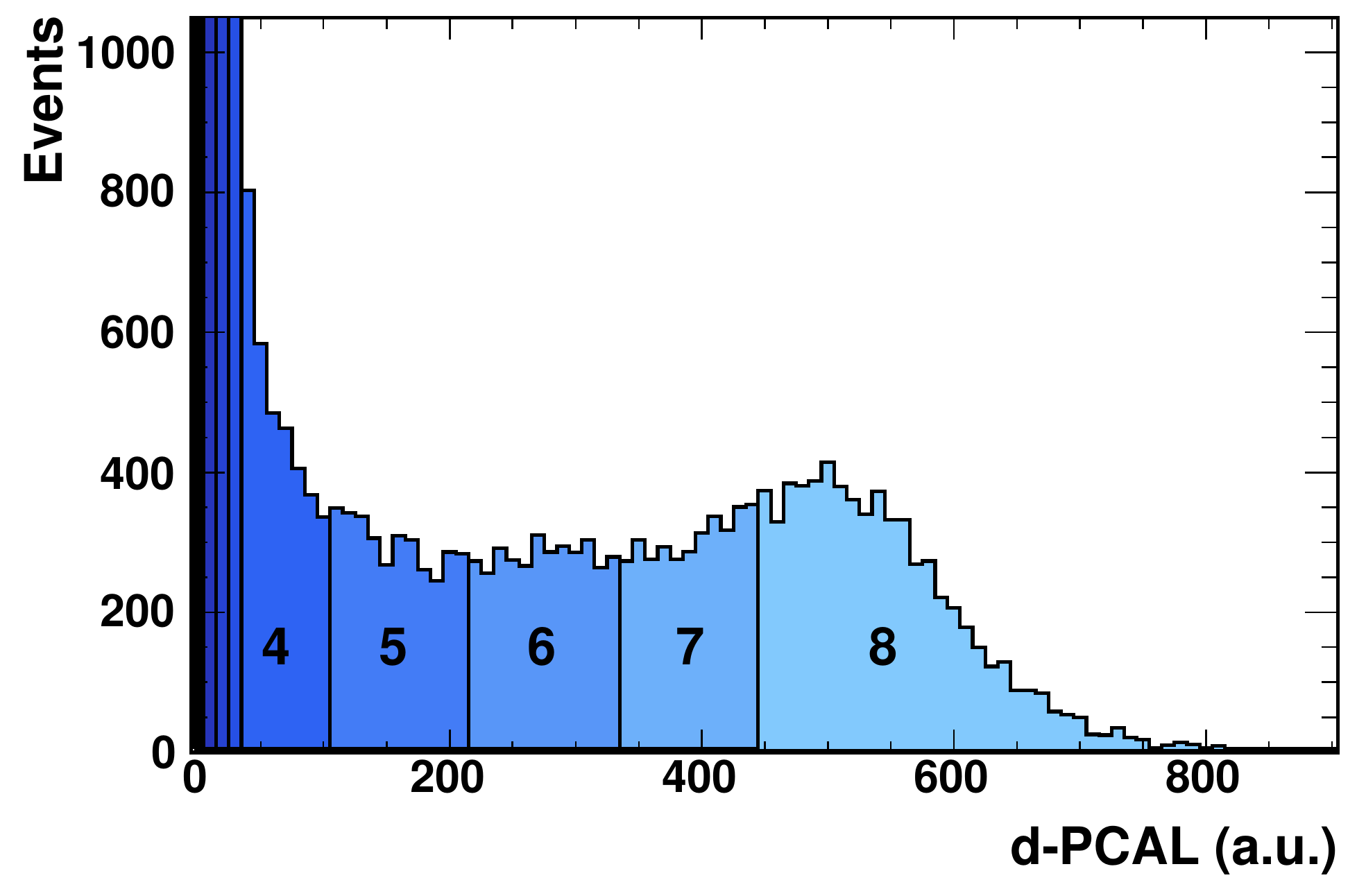}
   \caption{   \label{recon:fig:dPcalRegions}
      Regions used to study the characteristics of events with different total charge deposited in the d-PCAL (color online). Region 0 is the black colored bin located at the lowest detected d-PCAL signal.}
\end{figure}

\begin{figure}[th]
   \centering
   \includegraphics[width=0.95\linewidth]{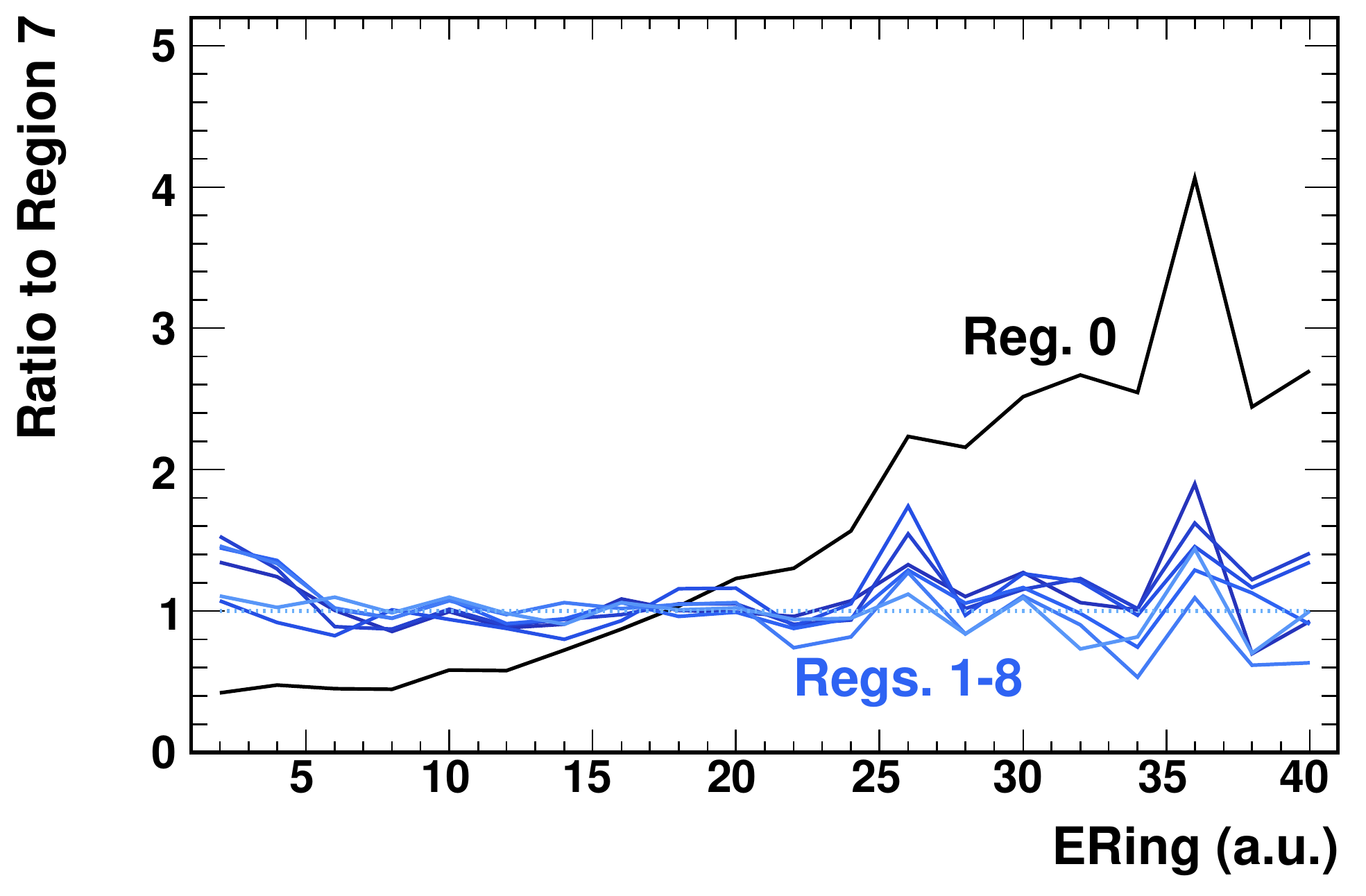}
   \caption{   \label{recon:fig:dPcalRatios}
     The ratio of the ERing distribution for events in each region of
     \fig{recon:fig:dPcalRegions} to that of region~7, which is
     partially under the proton peak. The color of the lines follows
     the same scheme as used in \fig{recon:fig:dPcalRegions}. Region
     0, in which no energy is deposited into the d-PCAL, shows a bias
     toward higher values of ERing, which is associated with more
     central collisions.}
\end{figure}

The observation of a spectator by one of the detectors is established
from the amount of energy deposited in that calorimeter. Because the
response of these calorimeters has not been simulated in the {\phob}
{\dAu} MC, the efficiency and purity of the chosen signal cuts cannot
be studied directly. Instead, the effect of the cuts on an independent
centrality measure (ERing) has been explored. This alternative method
is motivated by the expectation that tagging nucleon-nucleus
collisions should produce a data set that is biased toward
interactions with larger impact parameters than the full {\dAu} data
set.

The distribution of energy deposited in the d-PCAL is shown in
\fig{recon:fig:dPcalRegions} which has been divided into an arbitrary
set of regions numbered 0--8. While regions 7 and 8 show evidence of a
proton peak in the d-PCAL, events from all regions with non-zero
energy deposition (regions 1--8) show similar centrality
characteristics, as will be discussed below. Only events in region 0
show a bias toward more central collisions and are therefore assumed
to completely lack a proton spectator.

The presence or absence of a centrality bias in the regions displayed in \fig{recon:fig:dPcalRegions} is seen in \fig{recon:fig:dPcalRatios}, which shows the variation
in the shape of the ERing distribution for events depositing different
amounts of charge in the d-PCAL. Each line represents the ratio
between a particular region of \fig{recon:fig:dPcalRegions} and region
7. Collisions that deposit no energy into the d-PCAL show a striking
bias towards more central (higher ERing) interactions. Collisions in
regions 1--8 all show similar ERing distributions. This suggests that
any amount of energy deposited into the d-PCAL is due to a proton
spectator from the deuteron.

Furthermore, the observation that the shape of the ERing distribution
is the same for all collisions which deposit energy in the d-PCAL
supports the idea that these collisions are all of the same type,
namely {\nAu}.  As expected, the centrality of {\dAu} and tagged
{\nAu} collisions differ, but the centrality of {\nAu} does not depend
on the amount of energy that the spectator {\it proton} deposits in
the calorimeter.

\begin{figure}[th]
   \begin{center}
      \includegraphics[width=0.95\linewidth]{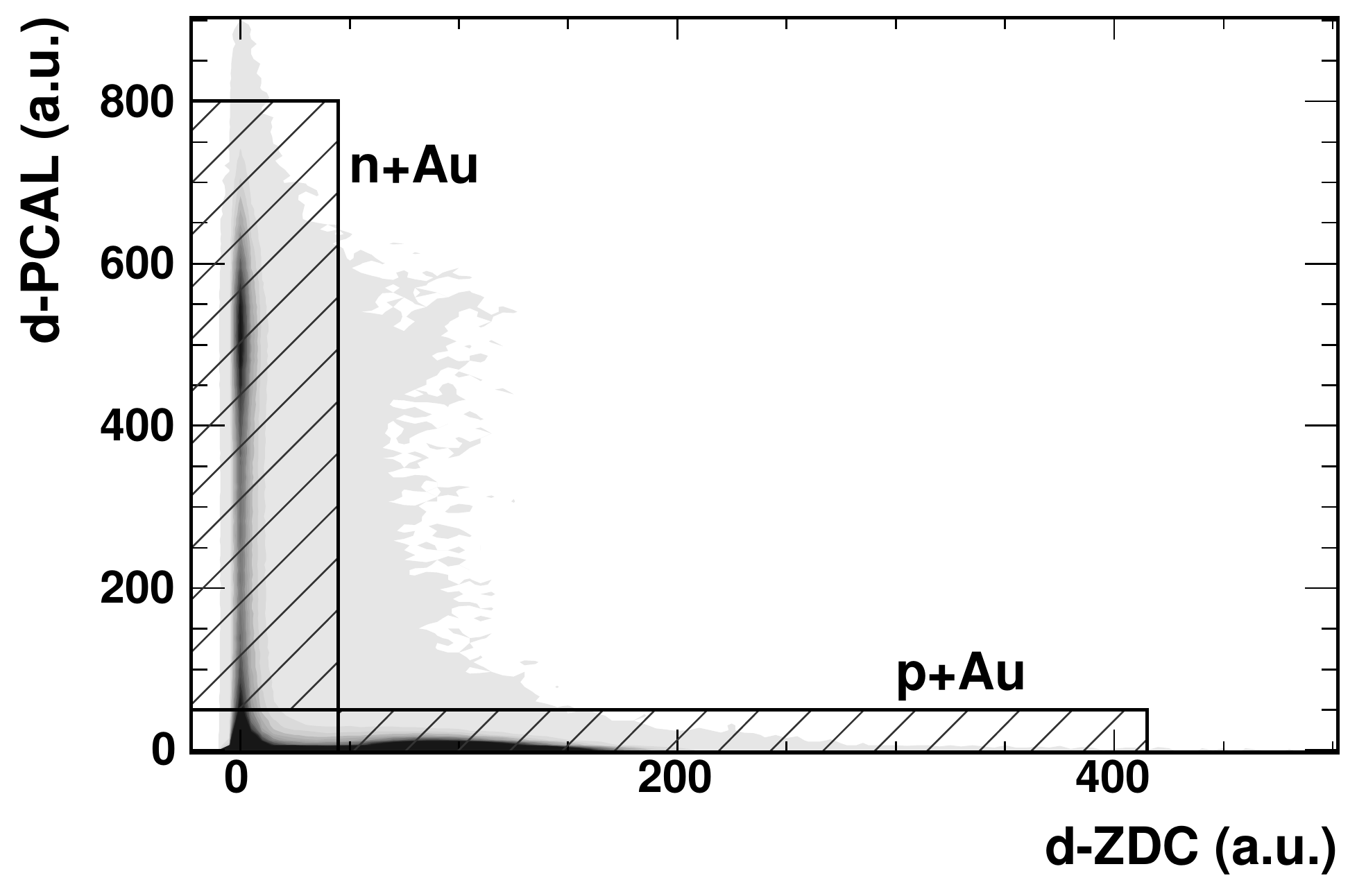}
   \end{center}
   \caption{   \label{recon:fig:dPcalVsdZDCRegs}
      The d-PCAL signal versus the d-ZDC signal. The boxes (at high
      d-PCAL, low d-ZDC and vice-versa) show the regions in which
      collisions can be identified as {\pAu} or {\nAu}.
      Note that the quadrant near (0,0) is not used to identify pure {\dAu}
      collisions, since it also contains nucleon-nucleus collisions
      (as the calorimeters are not perfectly efficient).}
\end{figure}

A similar procedure has been followed in order to determine the range
of energy deposited in the d-ZDC that corresponds to a neutron
spectator.  The final regions in which {\nAu} and {\pAu} interactions
are identified is shown in \fig{recon:fig:dPcalVsdZDCRegs}. The
minimum value of energy deposition in the d-PCAL is well above region
0, but ensures a very clean {\nAu} sample.

\subsection{Centrality of Nucleon-Nucleus Collisions}\label{coll:tag:cent}

The centrality of the tagged {\pAu} and {\nAu} collision data sets are
quantified (by parameters such as {\Ncoll}) within the fractional
cross section bins determined for {\dAu}. This is necessary because
the forward calorimeters are not included in the simulations of the
detector response, which precludes an event tagging procedure based on
the simulated energy deposition of those detectors.

Within a {\dAu} fractional cross section bin in the MC, the centrality
parameters of tagged events, such as {\Npart} in {\pAu} collisions,
are obtained using the true subset of simulated {\dAu} events
identified as {\pAu} or {\nAu}. These subsets are identified by the
presence of a neutron or proton, respectively, emerging from the
collision with a longitudinal momentum of 100~{\mom}.

The use of a tagging procedure based on true simulated momenta is
valid under the assumption that the event tagging procedure used in
data has an efficiency that does not depend on the centrality of the
collision.  That is, the average value of {\Ncoll} in {\pAu} is the
same whether the tagging efficiency is 80\% or 100\%, as long as the
tagging procedure does not alter the shape of the {\Ncoll}
distribution (but merely scales its normalization).

The validity of this assumption rests on three reasonable conjectures.
First, that it is not possible for a nucleon to both interact
inelastically and to still deposit a measurable amount of energy into
a forward calorimeter. Note that the Au-PCAL acceptance covers only
protons having no transverse momentum and a longitudinal momentum
$\abs{\pz}\gtrsim20~{\mom}$, and that the smaller d-PCAL will observe
only protons with even higher momenta. Second, that if a spectator
nucleon is present, it will be detected by a forward calorimeter with
an efficiency that is independent of the collision
centrality. Finally, that the forward calorimeter on the deuteron side
detects only deuteron spectators and not produced particles.

These conjectures imply that the tagging efficiency is independent of
the centrality of the {\dAu} collisions.  This allows nucleon-nucleus
collisions to be extracted from the {\dAu} (AMPT) simulations,
analogous to the tagging procedure used for data. Simulated nucleons
emerging from the interaction at the nucleon beam energy are taken to
be spectators. The centrality parameters extracted using this method
are presented in \Tab{raa:eringCentPars}.

The {\pAu} and {\nAu} events from the simulations have been used to
obtain a rough estimate of the tagging efficiencies. Taking the ratio
of {\pAu} ({\nAu}) to {\dAu} collisions that passed the event
selection in the simulation gives the fraction of {\pAu} ({\nAu})
events in the {\dAu} sample that would be tagged with a perfectly
efficient detector. Dividing the actual ratio of tagged {\pAu}
({\nAu}) to {\dAu} events found in the data by the fraction expected
from simulation gives an estimate of the efficiency. It is found that
\mbox{$\sim63\%$} of {\pAu} interactions and \mbox{$\sim46\%$} of
     {\nAu} interactions are tagged using the procedure described
     above. The lower {\nAu} efficiency may be at least partly due to
     the relatively large minimum d-PCAL energy required in the
     tagging procedure (see \sect{coll:tag}).


\section{Hadron Spectra Extraction}\label{spec}

The transverse momentum spectra of charged hadrons have been extracted
from tracks reconstructed using hits in the 16 layers of silicon
detectors that make up the two-arm magnetic spectrometer. Hit position
information is obtained both inside and outside the 2~T magnetic
field. Details of the vertex determination and particle tracking, as
used in previous {\phob} {\dAu} hadron spectra analyses, have been
described in Refs.~\cite*{Back:2003ns, Back:2003qr}. However, the
current studies make use of an expanded set of data and an updated
reconstruction procedure. As the {\dAu} collision trigger (described
in \sect{coll:trig}) does not include a high $\pt$ particle trigger,
as employed in~\cite{Back:2003ns}, a less biased data sample has been
used in the present analysis. To improve the efficiency of the
particle reconstruction, the final minimization step of the tracking
has been performed multiple times for each track. This helps prevent
the reconstruction from falling into a local minimum, which reduces
the number of both poor-quality track fits as well as ghost tracks.

In an effort to more accurately apply acceptance and efficiency
corrections, several changes have been made to the procedure used to
extract the hadron momentum spectra described in
Ref.~\cite{Back:2003ns}.  First, the geometrical acceptance and
tracking efficiency correction have been applied separately for each
of the two spectrometer arms. To account for acceptance effects as
accurately as possible, the correction factors as a function of $\pt$
have been applied as interpolated spline functions of the
track-embedding results (described in Ref.~\cite{Back:2003qr}), rather
than as smooth analytic functions.  Further, the minimum $\pt$ of
acceptable tracks has been lowered to correspond to the $\pt$ value at
which the acceptance and efficiency corrections are roughly 30\% of
their maximal value. This leads to a minimum $\pt$ value of 0.3 to
0.4~{\mom}, depending on the longitudinal collision vertex position,
for hadrons bending towards higher-$\eta$ (out of the {\phob}
spectrometer acceptance) and a minimum $\pt$ of about 0.1~{\mom} for
hadrons bending towards negative $\eta$.  Corrections for dead and hot
channels in the spectrometer have also been applied independently for
each spectrometer arm, to account for discrepancies on the level of
1\% in the hot and dead channel fraction of the two arms. The number
of ghost and secondary tracks passing the reconstruction cuts are
corrected for as a function of $\pt$. Due to improvements in the
reconstruction software since the publication of
Ref.~\cite{Back:2003ns}, these corrections are on the order of
1\%. Finally, corrections have been applied for the momentum
resolution of the tracking and the variable bin sizes of the
spectra. These corrections are determined using a dedicated simulation
of single particles through each spectrometer arm to determine the
distribution of reconstructed transverse momentum in each (true) $\pt$
bin.

The efficiency of the event selection described in \sect{coll:trig}
depends on centrality, particularly for peripheral events. Spectra uncorrected
for this effect would correspond to an ensemble of events with
a biased (higher) number of participants, rather than to a minimum bias selection using the same centrality binning. Instead, the efficiency determined as a
function of centrality (see \fig{ana:fig:ERingEfficiencies}) is used
to correct the spectra.

\begin{figure*}[t!]
   \begin{center}
      \includegraphics[width=0.95\linewidth]{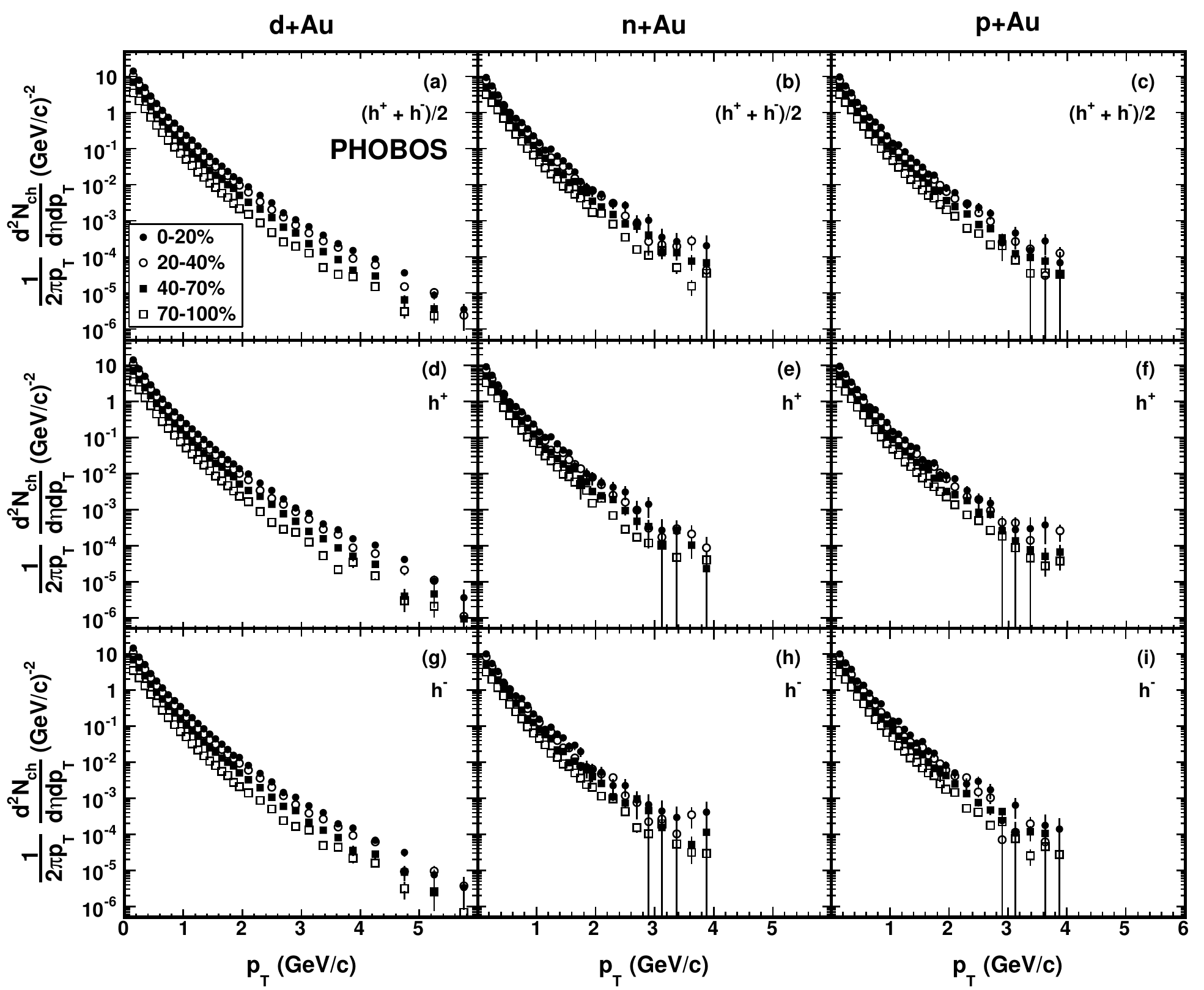}
   \end{center}
   \caption{\label{data:fig:ntSpecERingAMPT}
      The invariant yield of $\have$, $\hpos$, and $\hneg$ in four
      centrality bins determined for {\dAu} using the ERing
      centrality variable. The spectra for {\dAu}, {\nAu}, and {\pAu} are
      shown in separate columns. Due to the use of identical ERing
      cuts in all cases, the
      different data sets do not correspond to the listed fraction of
      the total inelastic
      cross section for nucleon-gold interactions. See text for details. Only statistical errors are shown.
      The spectra are obtained using particles that have a {\prap}
      $0.2<\eta<1.4$.}
\end{figure*}

The spectra of charged hadrons for {\dAu}, {\nAu} and {\pAu}
collisions are presented in \fig{data:fig:ntSpecERingAMPT} in four
bins of {\dAu} centrality, as determined by the ERing variable. 
For {\nAu} and {\pAu}, the same ERing cuts were used as for {\dAu}.
Therefore, these do not correspond to bins of the listed fractional
cross-section for nucleon-gold interactions.  Note that the
difference in the {\pt} range between {\dAu} and the nucleon-nucleus
spectra is simply due to fewer {\pAu} and {\nAu} collisions being
collected compared to {\dAu}.

\begin{figure}[th]
   \centering
   \includegraphics[width=0.95\linewidth]{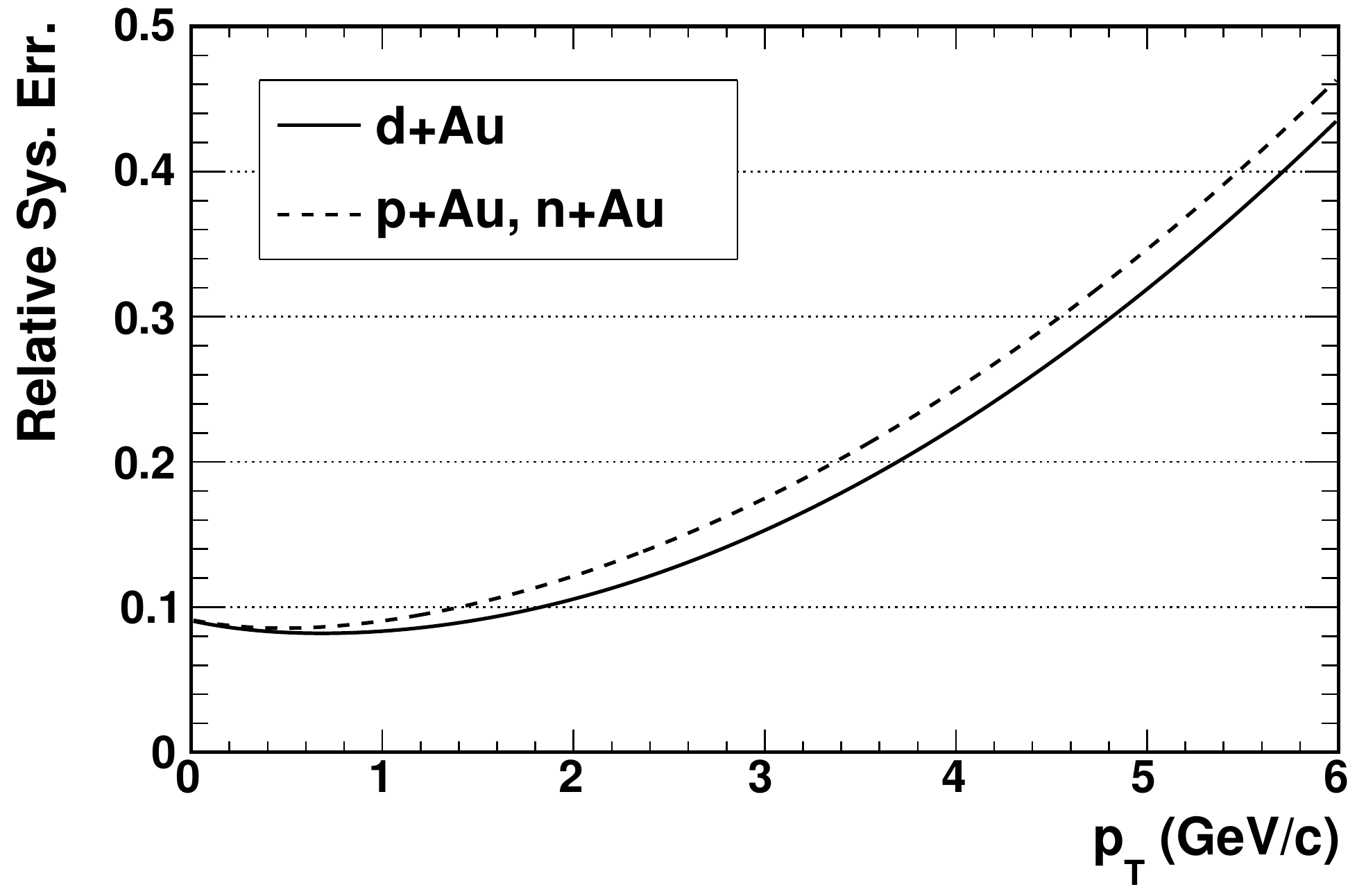}
   \caption{   \label{data:fig:syserrs}
      Contributions to the relative systematic uncertainty associated with uncertainty
      in the hadron spectra corrections.}
\end{figure}


Systematic uncertainties on the measured charged hadron spectra have
been quantified using the data. The largest correction, the acceptance
and efficiency of the tracking, is the source of the largest
systematic error (about 8\% at $\pt=2~\mom$). This error has been
estimated by comparing the yield in different subsets of the data for
which the particle spectrum is expected to be the same. For example,
the charged hadron yield of data taken with the spectrometer magnet in
the positive polarity is compared to that of data taken with the
magnet in the negative polarity. Similarly, yields measured separately
in each spectrometer arm have been compared in order to derive
uncertainties arising from the dead and hot channel correction. With
these corrections applied separately to each arm, the systematic
uncertainty on this effect is reduced to \mbox{$\lesssim3\%$} from
\mbox{$\sim10\%$} in the previous analysis~\cite{Back:2003ns}.

For corrections in which such studies are not possible, the
uncertainties are taken to be of the same order as the corrections
themselves. At $\pt=2~\mom$, this gives a ghost track uncertainty of
1\%, an uncertainty on the effect of secondary tracks of 3\% and an
uncertainty on the momentum resolution and momentum binning correction
(which are applied together) of about 3.5\%.

Uncertainty on the yield of nucleon-nucleus collisions due to tagging
has been estimated. This is done by varying the d-PCAL and d-ZDC cuts
used to tag events, which is expected to impact the number of
interactions in the data set, but not the yield of those
interactions. The total systematic uncertainties for the charged
hadron spectra are shown in \fig{data:fig:syserrs}.

\begin{figure}[t]
   \begin{center}
      \includegraphics[width=0.95\linewidth]{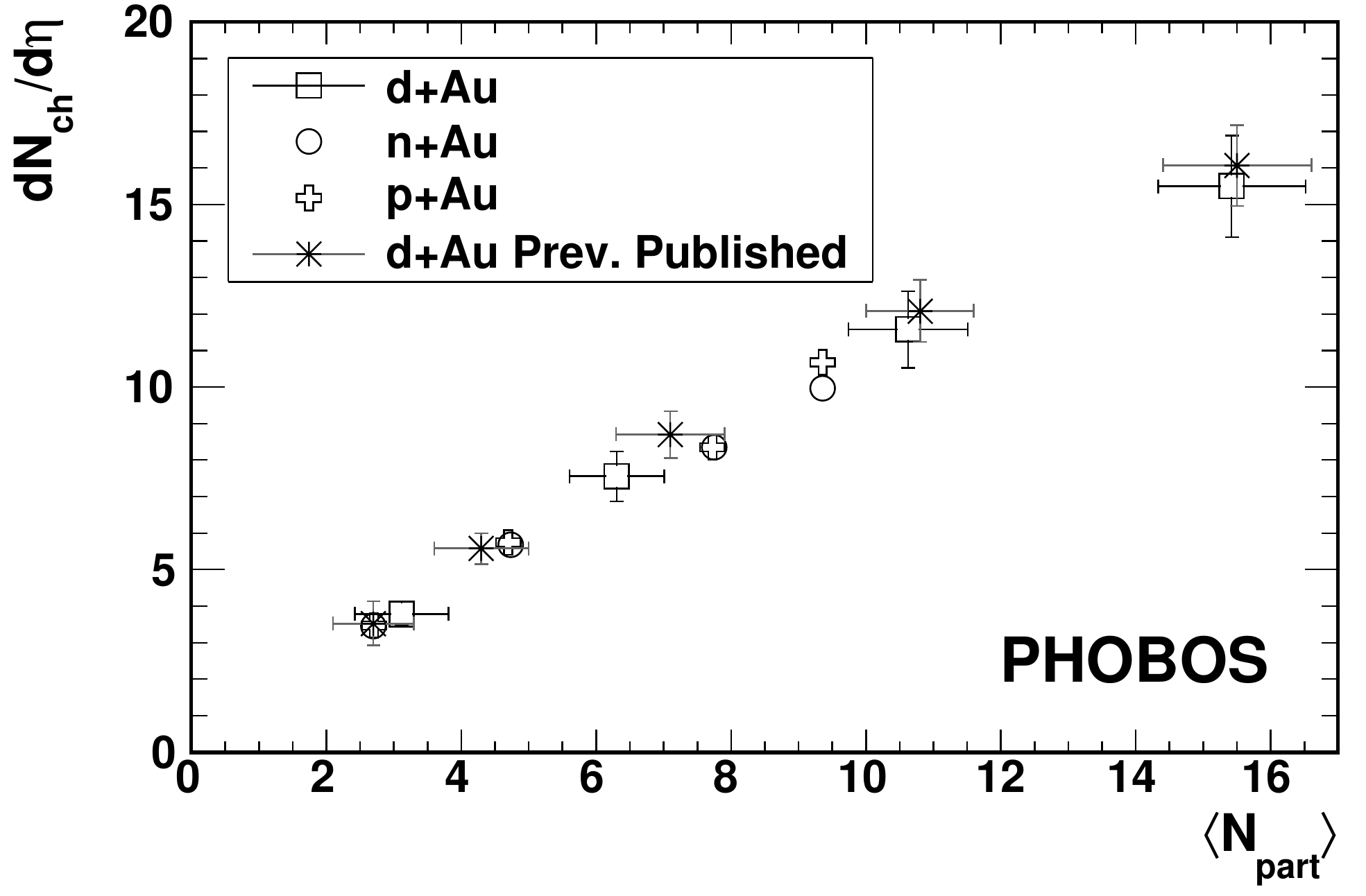}
   \end{center}
   \caption{\label{spec:eringTagMult} 
     The measured $\dnchdeta$ at $\aveeta=0.8$ in different collision
     systems obtained using ERing centrality bins. Systematic errors
     are shown as error bars for the {\dAu} measurements; statistical errors are
     negligible. Systematic errors on the nucleon-nucleus measurements
     are not shown, but are of similar order. Asterisk symbols show {\phob}
     multiplicities at {\mrap} from Ref.~\cite{phobWhitePaper, Back:BigMult}.}
\end{figure}


The charged hadron spectra are used to derive the multiplicity near
{\mrap} for {\dAu}, {\pAu} and {\nAu}. Spectra are modeled with
the following functional form
\begin{multline}
   \label{data:eq:rawfit}
\frac{1}{2 \pi \pt} \frac{d^{2}\!N_{\mathrm{ch}}}{\mathit{d\pt} \mathit{d\eta}} 
   = A \prn{1 + \frac{\pt}{p_0}}^{-n} + \\
     B \exp\prn{\frac{-\sqrt{\pt^2 + \mpi^2}}{T}}
\end{multline}
\noindent%
In the actual fit, parameter $A$ in Eq.~\ref{data:eq:rawfit} is
replaced by its value in terms of the analytically integrated yield
$\dnchdeta$ and the other four parameters.

\begin{multline}
  \label{data:eq:fitAparam}
A = \frac{(n-1) (n-2)}{2 \pi p_0^2} \Bigl[ \dnchdeta \: - \Bigr.\\
  \Bigl. 2 \pi B T (\mpi+T) e^{-\mpi / T} \Bigr]
\end{multline}
\noindent%
This allows both the value of $\dnchdeta$ and its statistical
uncertainty to be obtained directly from the fit. Systematic
uncertainties on the multiplicity are obtained by simultaneously
shifting each point in the spectra to the limit of its individual
systematic error and extracting $\dnchdeta$. The resulting systematic
uncertainty on the integrated yield is about 9\%.

The charged particle multiplicity near {\mrap}, at $\aveeta=0.8$, is
shown in \fig{spec:eringTagMult} for {\dAu}, {\pAu} and {\nAu} as a
function of {\Npart}. The number of participants is determined using
ERing centrality bins, since the ERing measurement of particles far
from {\mrap} has been shown to introduce at most a minimal bias on the
measurement \cite{phobWhitePaper}. A consistent dependence of the
multiplicity on {\Npart} is observed across all three collision
systems.

\section{An Improved Reference System}\label{raa}

The yield of hadrons in {\dAu} collisions has played a vital role in the
investigation of particle production in high energy {\AuAu}
collisions. The nuclear modification factor,
$R_X$, of a collision system, $X$, given by
\begin{equation}
   \label{rslt:eq:nucmod}
R_X = \frac{d^{2}\!N_X / \mathit{d\pt} \mathit{d\eta}}{%
   \avencl d^{2}\!N_{\mathit{p\bar{p}}} / \mathit{d\pt} \mathit{d\eta}}
\end{equation}\noindent%
where $X\!=${\AuAu}, {\dAu}, etc., has been used to test the scaling
of the high-$\pt$ hadron yield with the number of binary nucleon
interactions occuring during the collision. The nuclear modification
factor of nucleus-nucleus collisions at RHIC has been studied
extensively for {\AuAu} interactions at $\snn =
39~\gev$~\cite{Adare:2012uk} $62.4~\gev$~\cite{Back:2004ra,
  Arsene:2006pn, Adare:2012uk}, $130~\gev$~\cite{Adcox:2001jp,
  Adcox:2002pe, Adler:2002xw} and $200~\gev$~\cite{Back:2003qr,
  Adams:2003kv, Adler:2003qi, Adler:2003au, Abelev:2009wx,
  Adare:2012uk}, as well as for {\CuCu} interactions at $\snn =
22.4~\gev$~\cite{Adare:2008ad}, $62.4~\gev$~\cite{Adare:2008ad} and
$200~\gev$~\cite{Alver:2005nb, Adare:2008ad, Abelev:2009ab}.

One of the fundamental conclusions drawn from examination of the
nuclear modification factor is that the production of high-$\pt$ charged hadrons
in central {\AuAu} collisions at $\snn = 200~\gev$ is highly
suppressed with respect to binary collision
scaling~\cite{Back:2003qr}. However, it cannot be known from the
nucleus-nucleus data alone whether the suppression is due to
initial~\cite{Kharzeev:2002pc} or final~\cite{Baier:1996sk} state
effects. Nucleon-nucleus collisions at the same center of mass energy
would provide a control experiment capable of distinguishing between
the two possibilities, as such collisions should provide a nucleus in
the same initial state but should not produce an extended medium in
the final state. At RHIC these studies have been performed using
{\dAu} rather than nucleon-nucleus collisions~\cite{Back:2003ns,
  Adams:2003im, Adler:2003ii, Adler:2006xd, Arsene:2004ux,
  Arsene:2003yk}. The assumption was made
that, due to the small size and weak binding of the deuteron nucleus,
{\dAu} collisions would provide as good a control experiment for
{\AuAu} interactions as nucleon-nucleus collisions.

This assumption can be tested using tagged {\pAu} and {\nAu} collisions
to construct an improved reference for {\AuAu} collisions. Previous studies
performed by the NA49 collaboration~\cite{Fischer:2002qp,
Rybicki:2004jd} have suggested that hadron production of nucleus-nucleus
collisions may be better understood through careful consideration of the
neutron content of the nucleus. Taking into account the fact that a gold nucleus
consists of 60\% neutrons and 40\% protons, an improved  nuclear
modification factor for comparison to {\AuAu} can be defined as:
\begin{multline}
   \label{rslt:eq:RNAdef}
\RpnAu = 0.4 \frac{{(\dnchdetapA) / \avencl^{\mathit{pAu}}}}
{\dnchdetapp} + \\
   0.6 \frac{{(\dnchdetanA) / \avencl^{\mathit{nAu}}}}{\dnchdetapp}
\end{multline}
\noindent%
where $\avencl^{\mathit{pAu}}$ is the average number of collisions in
{\pAu}, $\avencl^{\mathit{nAu}}$ is the average number of collisions
in {\nAu}, and $\dnchdetapp$ is the yield of the reference
nucleon-nucleon system. 

The nucleon-nucleon reference comes from the UA1
measurement~\cite{Albajar:1989an} of the {\pbarp} inelastic cross
section. Note that data for {\pbarp} is used since data for the
preferable {\pp} system is not available at this energy. As described
in Ref.~\cite{Back:2003ns}, corrections are applied to the UA1 results
to account for (a)~the conversion from rapidity to {\prap} and (b)~the
difference between the UA1 acceptance (\mbox{$\abs{\smash[bt]{\eta}} <
  2.5$}) and the {\phob} acceptance (\mbox{$0.2 < \eta < 1.4$}). An
inelastic {\pbarp} cross section of 41~{\mb} is used to estimate the
yield of {\pbarp} collisions given the differential cross section
measurements from UA1.

\begin{figure*}
   \begin{center}
      \includegraphics[width=0.85\linewidth]{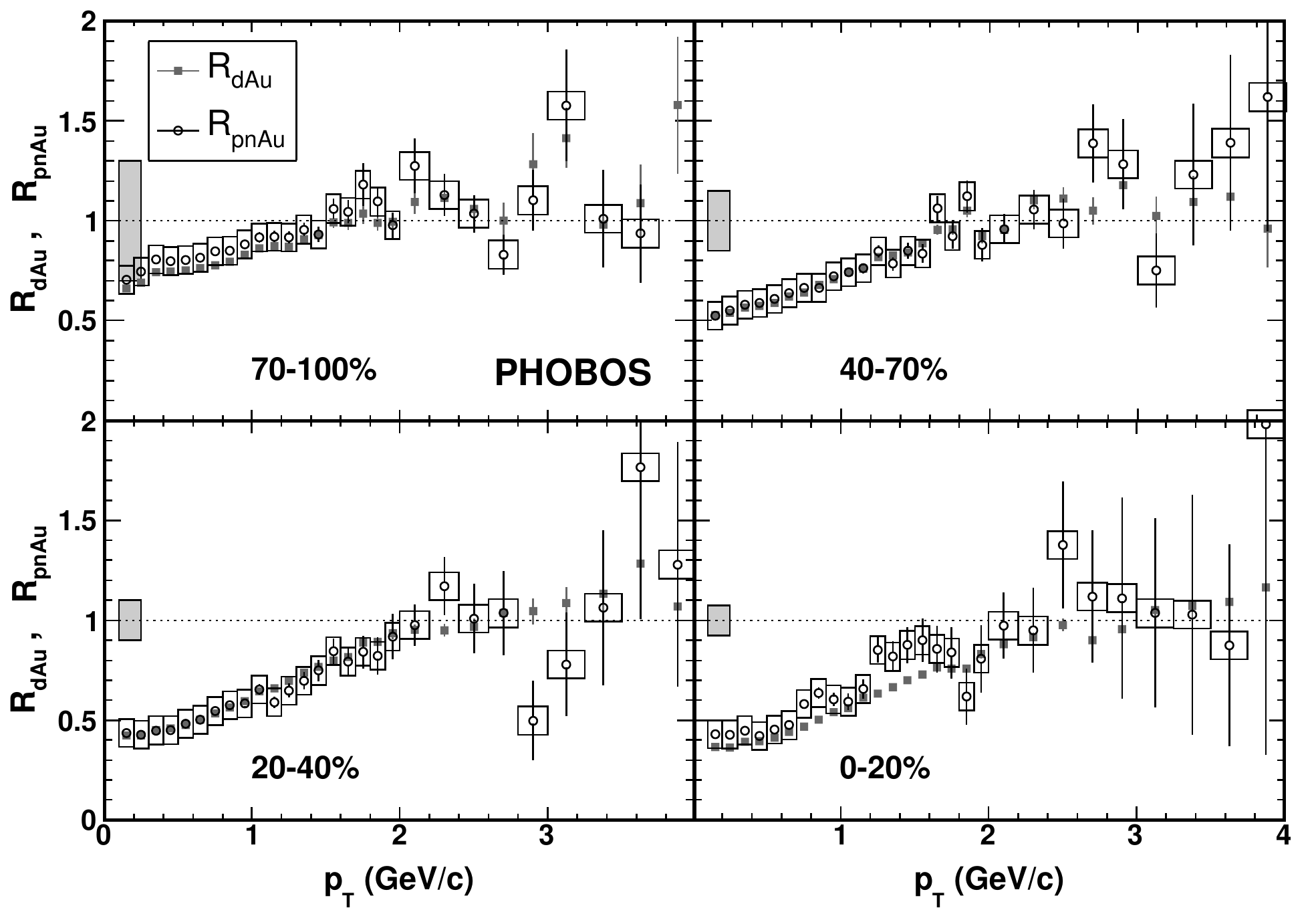}
   \end{center}
   \caption{\label{raa:RdAuRNAuCmp}
      Comparison of $\RdAu$ and $\RpnAu$ in each ERing centrality
      bin.  The height of the grey band shows the common scale
      uncertainty due to systematic errors on $\Ncoll$. The boxes
      around the $\RpnAu$ points show the supplemental systematic
      error on $\Ncoll$ in the {\NAu} system.}
\end{figure*}

The nuclear modification factor as a function of $\pt$ in the {\NAu} system, $\RpnAu$, is
compared to that of {\dAu}, $\RdAu$, for each centrality bin in \fig{raa:RdAuRNAuCmp}. Common systematic errors among the two
systems on the determination of $\Ncoll$ affect the overall scale of
the ratios, as shown by the height of the grey band. Further
systematic errors in the determination of $\Ncoll$ for the tagged
{\NAu} system are shown as boxes around the $\RpnAu$ points.

Qualitatively similar results have been found for a narrower window of
pseudorapidity by PHENIX~\cite{PHENIX_Tag}. The
$R_{\mathit{N}\mathrm{Au}}$ presented in that work is a simple average
of {\pAu} and {\nAu}, as opposed to the weighted combination shown in
\eq{rslt:eq:RNAdef}. While the shapes of the modification factors are
similar in this work and Ref.~\cite{PHENIX_Tag}, the latter appear to
be slightly shifted to larger values, most likely due to the use of
different reference spectra.

No significant difference between $\RpnAu$ and $\RdAu$ is
observed. This measurement supports the conclusions drawn from the
nuclear modification factor measurements of {\dAu}
collisions~\cite{Back:2003ns}; namely, that high-$\pt$ hadron
production in central {\AuAu} collisions is significantly suppressed
with respect to the expectation of binary collision scaling of
{\pbarp}~\cite{Back:2003qr}, while the production in {\dAu} collisions
is not. It should be noted that no claim of binary collision scaling
in {\dAu} or {\NAu} interactions has been made.



It has been observed that the nuclear modification factor in {\dAu}
exhibits a dependence on {\prap}~\cite{Back:2004bq, Arsene:2003yk,
  Arsene:2004ux, Arsene:2006pn}.  Thus, the apparent tendency of
$\RpnAu$ and $\RdAu$ to take the value of unity at high $\pt$ is
likely a consequence of the {\phob} {\prap} acceptance. Further, as
will be discussed in \sect{cronin}, the hadron production of {\dAu}
collisions is known to be \emph{enhanced} with respect to binary
collision scaling in a certain range of transverse momentum. Any
statement that {\dAu} lacks a suppression of high-$\pt$ hadrons is
therefore contingent upon the magnitude of this enhancement; see
Ref.~\cite{Accardi:2005fu} for a discussion.

Nevertheless, the stark discrepancy observed between {\NAu} and {\AuAu}
collisions at $\snn = 200~\gev$ demonstrate that final state effects
play a much stronger role in the high-$\pt$ hadron production of central
{\AuAu} collisions than do initial state effects. While the {\prap}
dependence of $\RdAu$ may provide evidence of some initial modification
of the gold nucleus~\cite{Kharzeev:2003wz, Jalilian-Marian:2005jf}, it
is clear that interactions with some dense, large volume medium produced
only in the nucleus-nucleus system forms the dominant source of
high-$\pt$ hadron suppression in {\AuAu} collisions. The data presented
here demonstrate that this conclusion is not biased by the use of
deuteron-nucleus rather than nucleon-nucleus interactions as the control
experiment for {\AuAu}.

\section{Centrality Dependence of the Spectral Shape}\label{cronin}

Although no clear evidence for enhancements above unity are seen in
the nuclear modification factor shown in \fig{raa:RdAuRNAuCmp}, the
{\pt} dependence may be related to the so-called Cronin effect. This
effect refers to the larger ratio of hadron production seen at high
{\pt} compared to lower {\pt} in proton-nucleus
collisions~\cite{Cronin:1974zm} relative to {\pp} collisions scaled by
the effective thickness of the nucleus.  General aspects of the
enhancement of inclusive charged hadron production (that is,
unidentified hadrons) in {\pAu} collisions can be described by models
in which partons undergo multiple scattering at the initial impact of
the {\pAu} collision~\cite{Accardi:2005fu}. However, the observed
difference in the strength of enhancement for mesons and
baryons~\cite{Shao:2006qi} is not easily explained by initial state
partonic scattering models. While other theories, such as those based
on the recombination model of hadronization~\cite{Hwa:2004zd}, may be
better suited to describe the enhancement of individual hadron
species, the shape of the {\dAu} $\pt$ spectrum relative to that of
$\pbarp$ is not a thoroughly understood phenomenon. Of particular
importance is the dependence of the spectral shape on the nuclear
thickness probed by the projectile (i.e.~the deuteron in a {\dAu}
collision)~\cite{Accardi:2003jh}.

The centrality dependence of the nuclear modification factor in {\dAu}
and {\AuAu} collisions at RHIC has been studied
extensively~\cite{phobWhitePaper, Adcox:2004mh, Adams:2005dq,
  Arsene:2004fa}. A particularly convenient method for exploring how
the shape of the transverse momentum spectra changes relative to
{\pbarp} has been suggested in Ref.~\cite{Back:2003ns}. This method
involves studying the centrality dependence of the charged hadron
yield in {\dAu} collisions relative to {\pp} at several values of
$\pt$. 



\begin{figure}[t!]
   \centering
   \includegraphics[width=0.95\linewidth]{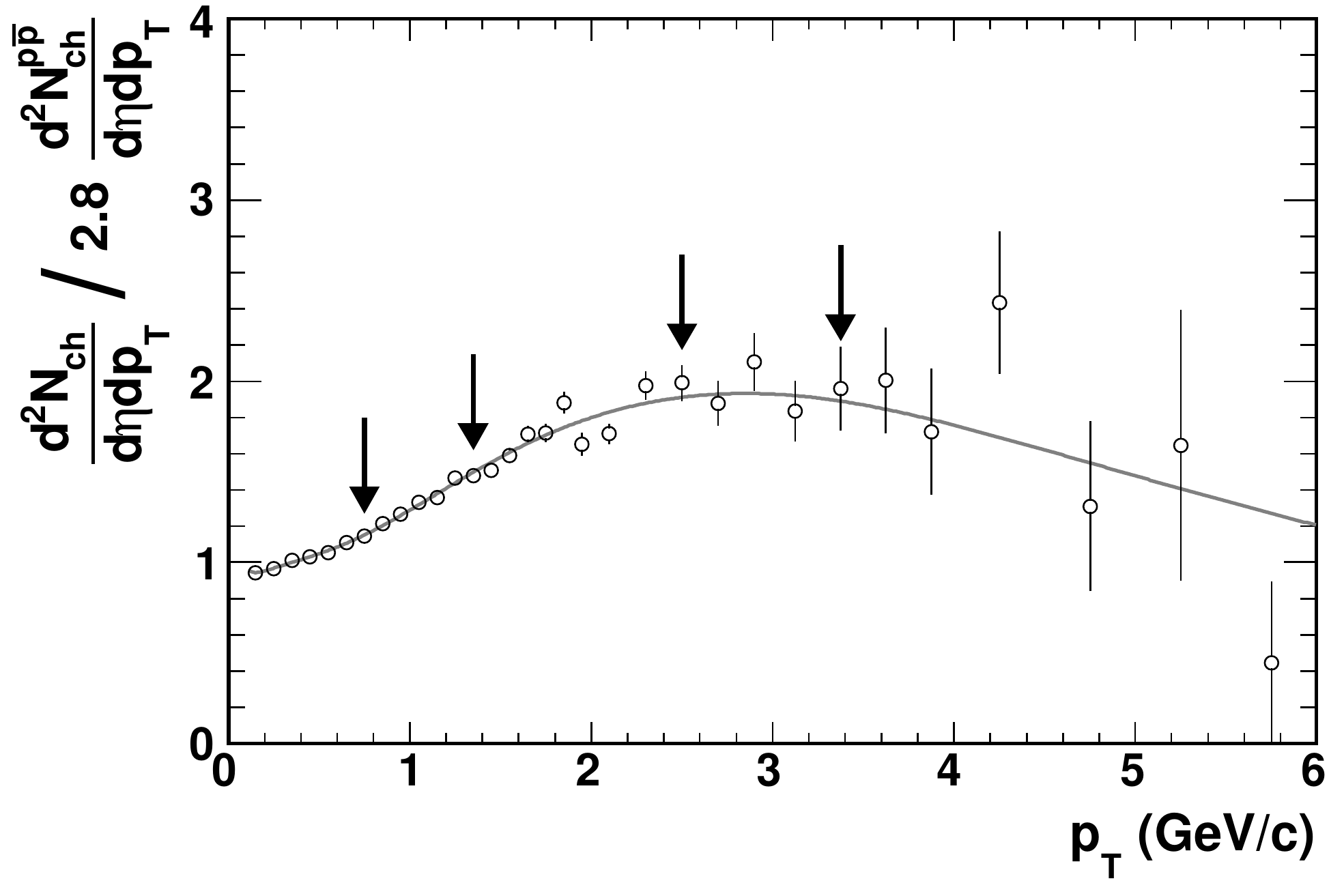}
   \caption{   \label{rslt:fig:exRlYldCronin}
     The ratio of ${d^{2}\!N_\mathit{ch}}/{\mathit{d\pt} \mathit{d\eta}}$ of {\dAu} in the 40-70\% ERing
     centrality bin to that in {\pbarp}, scaled by 2.8 so that
     the ratio is unity at $\pt=0.35$~{\mom} . The line is the ratio of fits to the two spectra using Eq.~\ref{data:eq:rawfit}. The arrows mark the
     $\pt$ values at which the centrality dependence of the relative
     yield is studied (see \fig{cronin:relYldvsdNdeta}).}
\end{figure}

The procedure for determining the so-called relative yield is
as follows. First, the transverse momentum spectrum in a particular {\dAu}
centrality bin is compared to the spectrum of {\pbarp}. To compare
only the shape of the two spectra, they are then normalized such that
the spectra match at $\pt = 0.35~\mom$. While this specific value of
$\pt$ is arbitrary, it has been intentionally chosen to be in a region
where soft processes drive particle production. Matching the {\dAu}
spectra to the {\pbarp} spectra serves to remove any trivial
enhancement of hadron production in {\dAu} that is simply due to the
larger number of nucleon-nucleon collisions occurring in that
system. However, matching in this way does not assume $\Ncoll$
scaling, nor does it have any effect on the relative shape of the
spectra.

Next, the ratio of the normalized {\dAu} spectra and the {\pbarp}
spectra is determined. The value of this ratio at certain transverse
momentum values are selected, as shown in
\fig{rslt:fig:exRlYldCronin}. Finally, the centrality dependence of
the normalized ratio, the relative yield, at the chosen $\pt$ values
is studied.

\begin{figure*}[t]
   \begin{center}
      \includegraphics[width=0.9\linewidth]{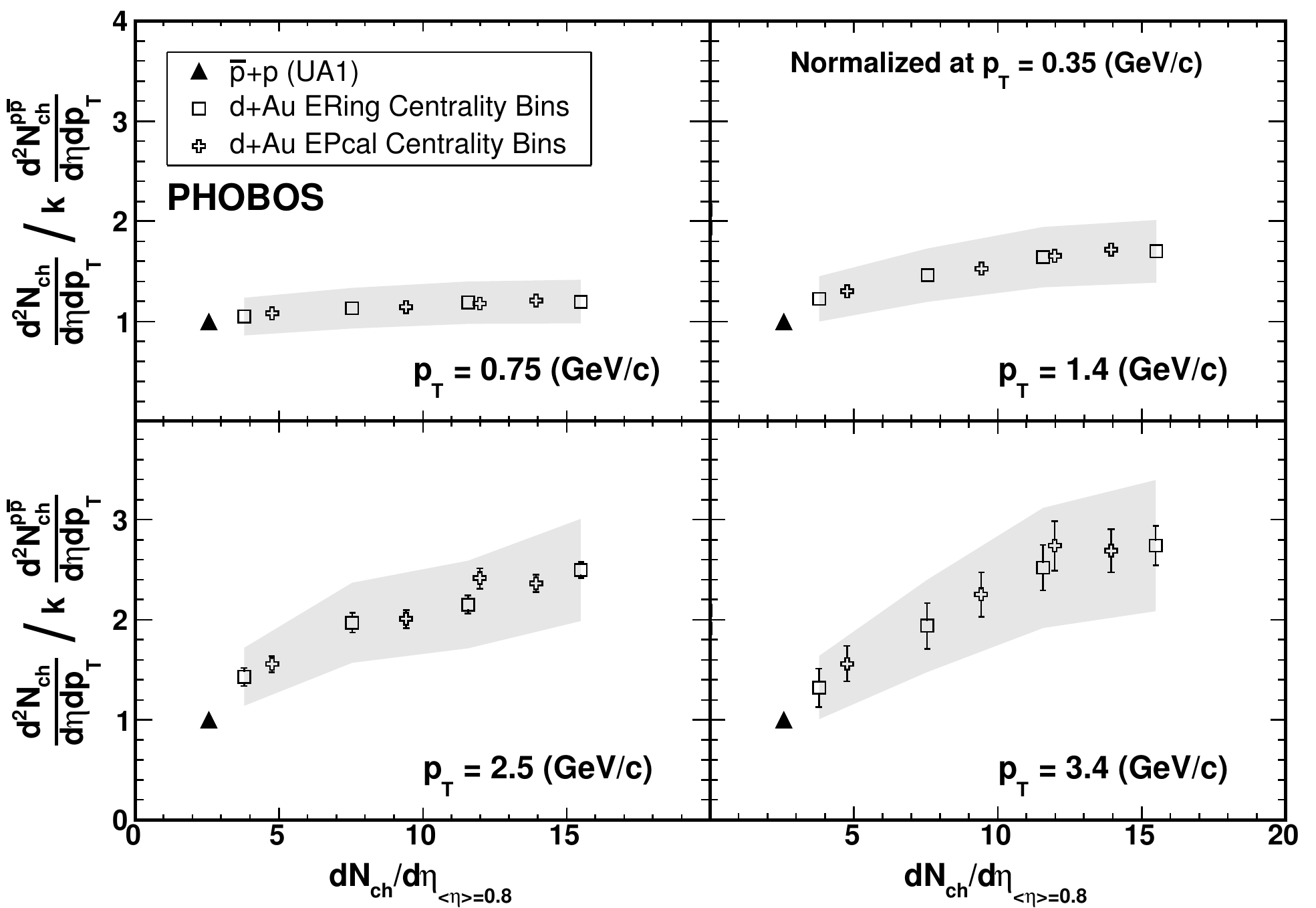}
   \end{center}
   \caption{\label{cronin:relYldvsdNdeta}
     Open symbols show the average hadron yield of {\dAu} collisions relative to
     {\pbarp} as a function of {\dAu} $\dnchdeta$ near {\mrap}, scaled by a
     factor, $k$, such that the ratio is unity at $p_{T}=0.35$~{\mom}
     in order to focus on the evolution of the shape of the
     yield. Statistical errors are represented by bars on the
     points. The systematic error for the ratio in ERing centrality
     bins is shown by the grey band. See text for a discussion of the
     systematic errors. Closed triangles at a relative yield of 1 (representing {\pbarp} divided by itself) are plotted at the 
    $\dnchdeta$ for {\pbarp}. The dependence of the relative yield on both
     centrality and $\pt$ is observed to  extrapolate smoothly back to
     {\pbarp}.}
\end{figure*}

The relative yield of {\dAu} collisions to {\pbarp} is shown in
\fig{cronin:relYldvsdNdeta} as a function of $\dnchdeta$, for four
different values of transverse momentum. It is expected that systematic
effects on the relative yield are highly correlated between the spectra
measured with different centrality bins. Thus, shifts in the relative
yield will tend to move all points together. See \tab{raa:eringCentPars}
for a description of the systematic uncertainties on the centrality
variables measured with ERing. With centrality parametrized by the
experimentally measured integrated yield, no bias or (Glauber) model
dependence is introduced by the choice of centrality variable.

From \fig{cronin:relYldvsdNdeta}, it is clear that the difference between the
{\dAu} and {\pbarp} spectra depends on both centrality and $\pt$. If the shape of
the two spectra were identical, the relative yield
would be constant at unity for all values of $\pt$ and centrality.
Instead, the {\dAu} spectra show an enhancement over {\pbarp} that
increases with centrality. The strength of this enhancement is observed
to increase at higher $\pt$. It would be interesting to study the
relative yield of much higher $\pt$ hadrons, on the order of 10 to
100~{\mom}, in order to test whether the shape of the {\pbarp} spectra
is recovered in hard scattering processes. However, such particles are
produced very rarely and too few are present in the {\phob} data set to
allow such a study.

Nevertheless, the data show a smooth extrapolation of the relative yield
of {\dAu} collisions to that of {\pbarp} as the {\dAu} collisions become
more peripheral. Thus, distortions of the {\dAu} spectra caused by
nuclear effects diminish in a smooth way as the amount of nuclear
material probed by the deuteron is reduced. 
The integrated charged particle yield near $\ave{\eta}\approx 0.8$ has been chosen as the centrality measure, since it provides a model-independent variable with which to study the centrality dependence of hadron production in nucleon-nucleus and nucleus-nucleus systems.


\section{Charge Transport}\label{asym}

The availability of both {\pAu} and {\nAu} collision data presents a
unique opportunity to study baryon transport in nucleon-nucleus
collisions. Since a {\pAu} collision contains one more charged hadron
than an {\nAu} collision, a search for this extra charge near the
{\mrap} region is possible. Previous measurements~\cite{Alber:1997sn}
of {\pAu} collisions at $\snn = 19.4~\gev$ found that the number of
net protons (p - \={p}) per unit of rapidity is less than one in the
{\mrap} region. In addition, studies have shown a decrease in the
{\mrap} net proton yield with increasing center of mass energy; see
Ref.~\cite{Back:2006tt} for a discussion. Further, it has been
inferred that hadrons traversing nuclear material do not lose more
than about two units of rapidity~\cite{Busza:1983rj}. Thus, it is
expected that any charge asymmetry between hadrons measured at {\mrap}
in {\pAu} and {\nAu} collisions would be small.

Nevertheless, a comparison of charged hadron production in {\pAu} and
{\nAu} allows the transport of charge explicitly from the projectile
proton to be studied. Assuming that baryons from the gold nucleus
undergo transport to {\mrap} via the same process in both {\pAu} and
{\nAu} collisions, any charge transport to {\mrap} of protons in the
gold nucleus would not lead to an asymmetry.

Simple charge conservation would imply that the \emph{total} number of
positive particles emerging from a {\pAu} collision should be greater
(by one) than the number emerging from a {\nAu} collision. Whether or
not this charge asymmetry is present near {\mrap} has been studied
using the observable $A^{\mathit{pn}}_{h^{\pm}}$, defined as
\begin{equation}
   \label{rslt:eq:spn}
A^{\mathit{pn}}_{h^{\pm}} = \frac{\dnhpndetapA - \dnhpndetanA}
{\dnhpndetapA + \dnhpndetanA}
\end{equation}
\noindent%
where $A^{\mathit{pn}}_{h^{+}}$ denotes the asymmetry between {\pAu}
and {\nAu} in the yield of positively charged hadrons at $\aveeta=0.8$
and $A^{\mathit{pn}}_{h^{-}}$ denotes the asymmetry of the yield of
negatively charged hadrons between the two systems.



\begin{figure}[th]
   \centering
   \includegraphics[width=0.95\linewidth]{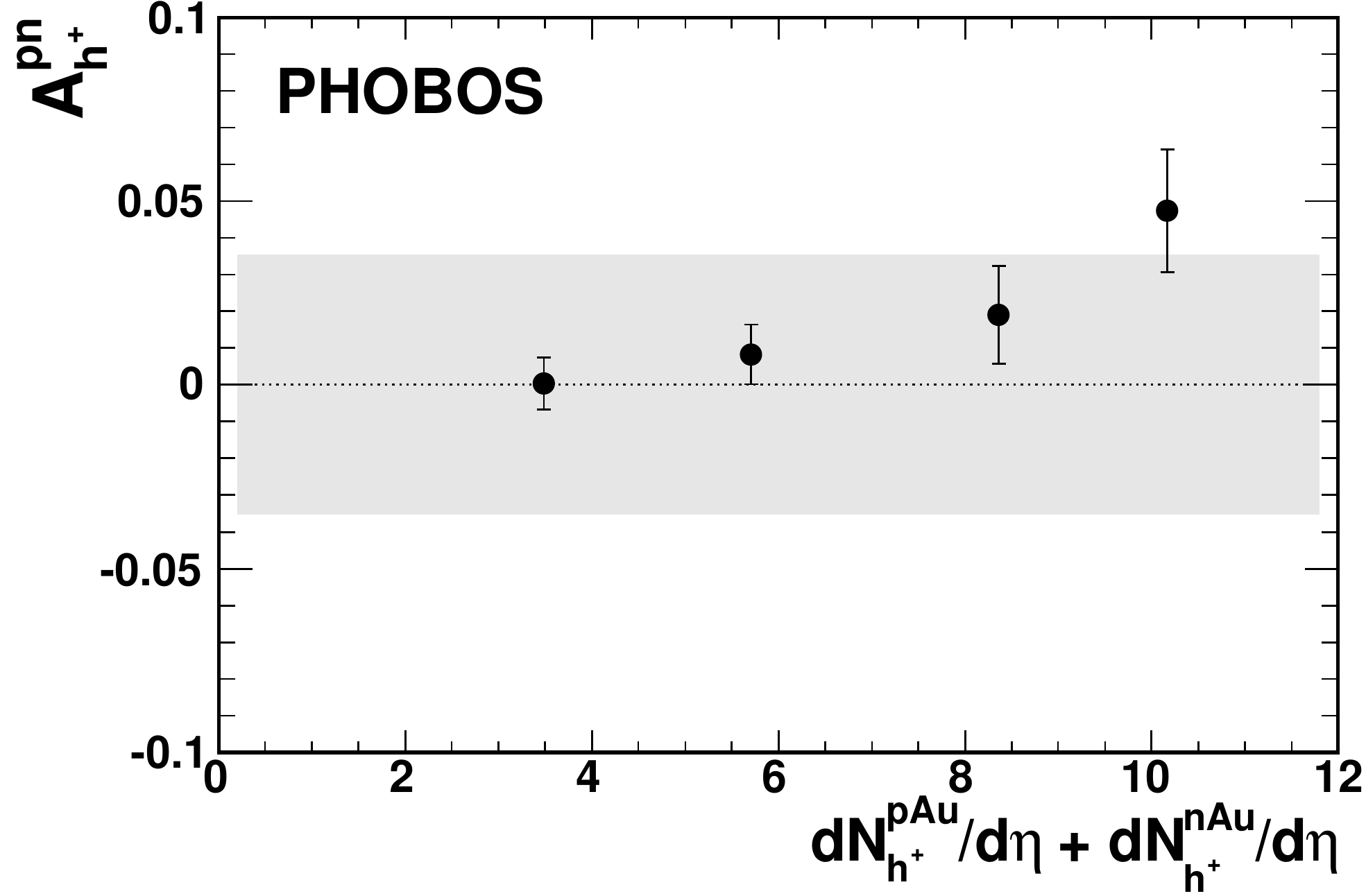}
   \caption{   \label{asym:asymmMultPos}
      The asymmetry of positive hadrons between {\pAu} and {\nAu}
      collisions at $\aveeta=0.8$ as a function of centrality. The
      grey band shows the systematic uncertainty in the overall scale
      of the ratio.}
\end{figure}

\begin{figure}[th]
   \centering
      \includegraphics[width=0.95\linewidth]{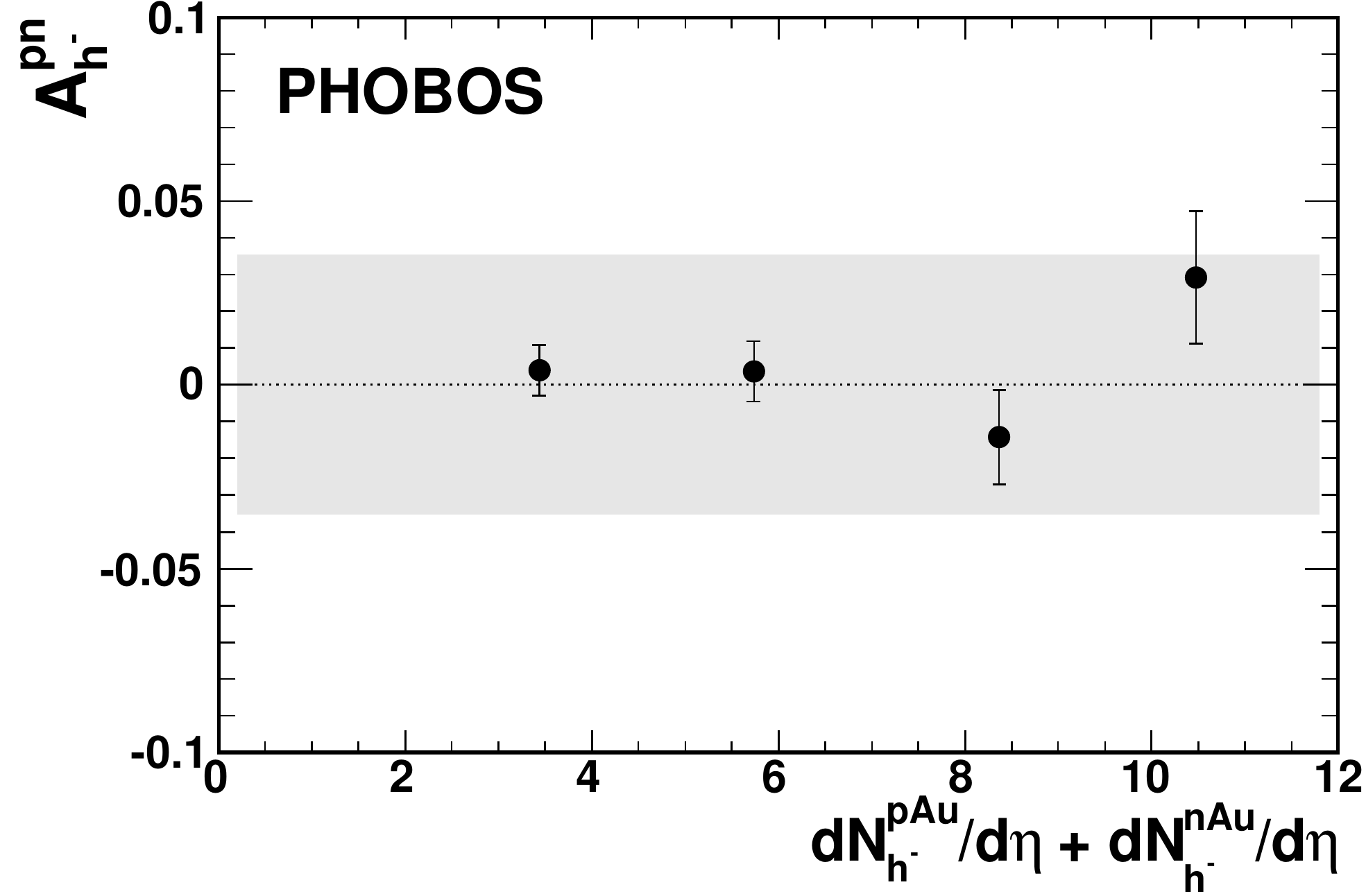}
   \caption{   \label{asym:asymmMultNeg}
      The asymmetry of negative hadrons between {\pAu} and {\nAu}
      collisions at $\aveeta=0.8$ as a function of centrality. The
      grey band shows the systematic uncertainty in the overall scale
      of the ratio.}
\end{figure}

The charge asymmetry defined by \eq{rslt:eq:spn} is presented in
\fig{asym:asymmMultPos} for positive hadrons and in
\fig{asym:asymmMultNeg} for negative hadrons. The grey band in each
figure represents the systematic uncertainty in the asymmetry ratio,
propagated from the nucleon tagging component of the systematic
uncertainty on the momentum spectra (see \sect{spec}).  Only
uncertainties specific to reconstructing the nucleon-nucleus $\pt$
spectra contribute to this systematic error, as all other effects
divide out in the ratio. No evidence for asymmetry between {\pAu} and
{\nAu} collisions is observed at $\aveeta = 0.8$, which is slightly
forward on the deuteron-going side.



\section{Summary}\label{summ}

The addition of two forward proton calorimeters to the {\phob}
detector allows the extraction of {\pAu} and {\nAu} collisions from
the {\dAu} data set. Centrality parameters have been determined for
each of the collision systems using observables based on the
multiplicity at high rapidity and on the number of spectators. The
number of particles produced near {\mrap} is found to scale with
{\Npart} across all collision systems. The charged hadron spectra have
been measured for {\pAu}, {\nAu}, and {\dAu} collisions and used to
construct an ideal nucleon-nucleus reference for {\AuAu}
collisions. The nuclear modification factor of this ideal reference is
found to agree with that of {\dAu}. The shape of the nuclear
modification factor has been studied in detail and is found to depend
on both centrality and transverse momentum. A larger ratio of the
{\dAu} over {\pbarp} spectra is found at larger values of {\pt} and
this enhancement is found to extrapolate smoothly as a function of
multiplicity at {\mrap} from {\pbarp} to central {\dAu} collisions.
Finally, a comparison of the yield of positively and negatively
charged hadrons in {\pAu} and {\nAu} has been conducted in a direct
search for evidence of charge transport to {\mrap}. No significant
asymmetry between the charged hadron yields in {\pAu} and {\nAu} is
observed at $\aveeta=0.8$.

%
%
%
%
This work was partially supported by U.S. DOE grants 
DE-AC02-98CH10886,
DE-FG02-93ER40802, 
DE-FG02-94ER40818,  
DE-FG02-94ER40865, 
DE-FG02-99ER41099, and
DE-AC02-06CH11357, by U.S. 
NSF grants 9603486, 
0072204,            
and 0245011,        
by Polish MNiSW grant N N202 282234 (2008-2010),
by NSC of Taiwan Contract NSC 89-2112-M-008-024, and
by Hungarian OTKA grant (F 049823).


\bibliography{NucTagPaper}

\end{document}